\documentclass[12pt]{iopart}
\usepackage{iopams}

\usepackage{graphicx}% Include figure files
\usepackage{dcolumn}% Align table columns on decimal point
\usepackage{bm}% bold math

\usepackage{units}
\usepackage{cite,mcite}
\usepackage{subfigure}
\usepackage{url,multirow}

\def\NeNmu{\mbox{$(N_{\rm e},N_{\mu})$}}

\begin{document}

\title{On the measurement of the proton-air cross section using air
  shower data}
\author{R~Ulrich\footnote{Corresponding author,
    ralf.ulrich@kit.edu}, J~Bl{\"u}mer, R~Engel, F~Sch{\"u}ssler, and
  M~Unger} \address{%
  Karlsruhe Institute of Technology (KIT)\footnote{KIT is the
    cooperation of
    University Karlsruhe and Forschungszentrum Karlsruhe},\\
  Institut f\"ur Kernphysik, P.O. Box 3640, 76021 Karlsruhe, Germany }
% \ead{ralf.ulrich@kit.edu}

\date{\today}% It is always \today, today,
% but any date may be explicitly specified

\submitto{NJP}

\begin{abstract}
  The analysis of high-energy air shower data allows one to study the
  proton-air cross section at energies beyond the reach of fixed
  target and collider experiments. The mean depth of the first
  interaction point and its fluctuations are a measure of the
  proton-air particle production cross section. Since the first
  interaction point in air cannot be measured directly, various
  methods have been developed in the past to estimate the depth of the
  first interaction from air shower observables in combination with
  simulations.  As the simulations depend on assumptions made for
  hadronic particle production at energies and phase space regions not
  accessible in accelerator experiments, the derived cross sections
  are subject to significant systematic uncertainties. The focus of
  this work is the development of an improved analysis technique that
  allows a significant reduction of the model dependence of the
  derived cross section at very high energy. Performing a detailed
  Monte Carlo study of the potential and the limitations of different
  measurement methods, we quantify the dependence of the measured
  cross section on the used hadronic interaction model. Based on these
  results, a general improvement to the analysis methods is proposed
  by introducing the actually derived cross section already in the
  simulation of reference showers. The reduction of the model
  dependence is demonstrated for one of the measurement methods. 
\end{abstract}

\maketitle

%\linenumbers

\section{Introduction}
The natural beam of cosmic ray particles extends to energies far
beyond the reach of any Earth-based particle accelerator. Therefore
cosmic ray data provide a unique window to study hadronic interaction
phenomena at energies up to several Joules per particle, corresponding
to an equivalent center-of-mass energy of up to $\unit[450]{TeV}$.

On the other hand, direct observation of the first interaction of
ultra-high energy cosmic rays in the upper atmosphere is impossible
due to the very low flux of these particles. Only the cascades of
secondary particles, called extensive air showers (EAS), can be
measured with arrays of particle detectors or optical telescopes. To
obtain information on the first interactions in an air shower it is
necessary to link the measured air shower characteristics to that of
high energy particle production in the shower. This can be done with
detailed Monte Carlo simulations of the shower evolution and the
corresponding shower observables, but inevitably causes a dependence
of the results on hadronic interaction models needed for the shower
description.

The total particle production cross section is one of the most
fundamental quantities that characterizes hadronic
interactions. Considering proton-induced air showers of the same
primary energy, the depth of the first interaction point, $X_1$, is
distributed according to
\begin{equation}
  \frac{{\rm d}p}{{\rm d}X_1}=\frac{1}{\lambda_{\rm p-air}}\, {\rm
    e}^{-X_1/\lambda_{\rm p-air}}\,,
\end{equation}
where $\lambda_{\rm p-air}$ is the interaction mean free path of
protons in air. The mean depth of the first interaction point and its
shower-to-shower fluctuations are directly linked to the size of the
proton-air cross section by
\begin{equation}
  \sigma_{\rm p-air} = \frac{\langle m_{\rm air}\rangle}{\lambda_{\rm p-air}} \,,
\end{equation}
with the mean target mass of air being $\langle m_{\rm
  air}\rangle\approx 14.45\,m_{\rm
  p}=\unit[24160]{mb\,g/cm^2}$~\cite{Bodhaine:1999}.  It is,
therefore, not surprising that there is a long history of attempts to
infer this cross section from high-energy cosmic ray
data~\cite{Grigorov:1965aa,Yodh:1972fv,Nam:1975aa,Siohan:1978zk,
  Mielke94,Hara:1983pa,Honda:1993kv,Nikolaev:1993mc,%
  Aglietta:2009zz,Baltrusaitis:1984ka,Knurenko:1999cr,Belov:2006mb,
  Collaboration:2009ca,Gaisser:1993ix}.
 
%%%%%%%%%%%%%%%%%%%%%%%%%%%%%%%
\begin{figure}[bt!]
  \centering
  \includegraphics[width=.9\linewidth]{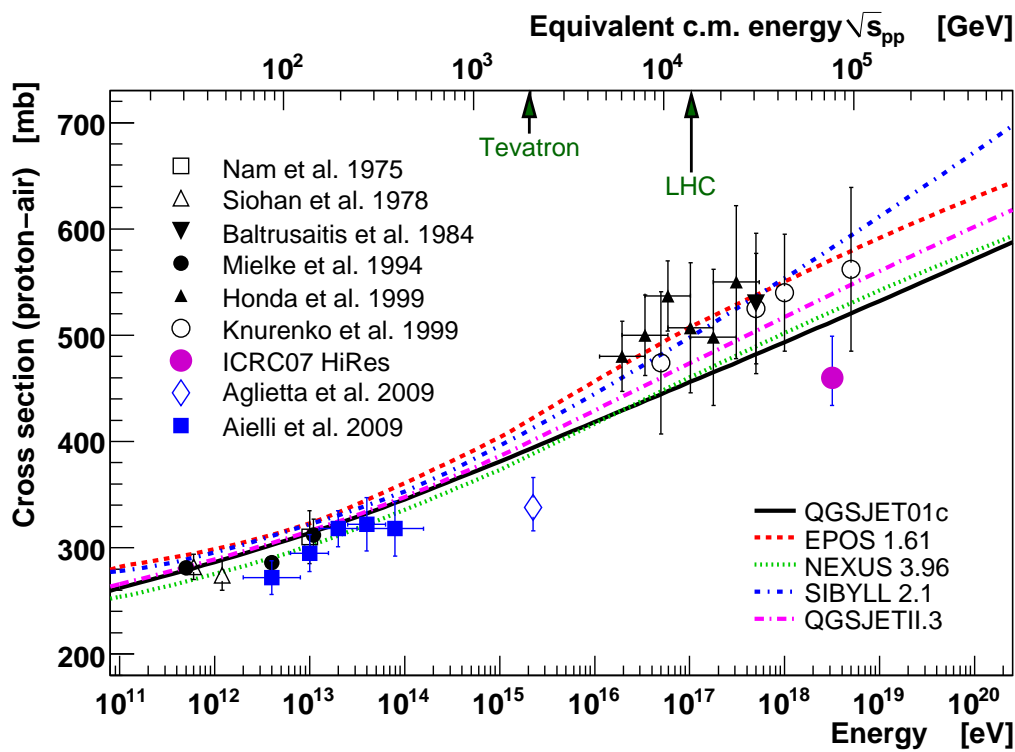}
  \caption{Compilation of proton-air
    production cross sections from cosmic ray
    measurements
    \protect\cite{Grigorov:1965aa,Yodh:1972fv,Nam:1975aa,Siohan:1978zk,%
      Mielke94,Hara:1983pa,Honda:1993kv,Nikolaev:1993mc,%
      Aglietta:2009zz,Baltrusaitis:1984ka,Knurenko:1999cr,Belov:2006mb,Collaboration:2009ca}.
    The data are compared to model predictions~\protect\cite{%
      Kalmykov:1997te,Ostapchenko:2005nj,Engel:1999db,Fletcher:1994bd,Werner:2007vd,%
      Ostapchenko:2004ss}.}
  \label{f:SigmaData}
\end{figure}
%%%%%%%%%%%%%%%%%%%%%%%%%%%%%%%

A compilation of published proton-air cross section measurements and
predictions of ultra-high energy hadronic interaction models is shown
in Fig.~\ref{f:SigmaData}. All data above \unit[100]{TeV} are based on
the analysis of EAS
data~\cite{Hara:1983pa,Honda:1993kv,Nikolaev:1993mc,%
  Aglietta:2009zz,Baltrusaitis:1984ka,Knurenko:1999cr,Belov:2006mb}.
Also the ARGO-YBJ measurements around \unit[10]{TeV} are originating
from a high altitude EAS array~\cite{Collaboration:2009ca}. The data
at lower energies stem from unaccompanied hadron
analyses~\cite{Grigorov:1965aa,Yodh:1972fv,Nam:1975aa,Siohan:1978zk,%
  Mielke94}.

All cosmic ray measurements of the proton-air cross section are only
sensitive to the \emph{particle production} cross
section~\cite{Nikolaev:1993mc,Engel:1998pw}.  In addition,
interactions with insignificant particle production have no measurable
impact on cosmic ray observations. This is the case for air shower
based techniques as well as for the unaccompanied hadron method.

To draw conclusions on the energy of the primary particle and its
first ultra-high energy interactions, all relevant processes involved
from the primary cosmic ray particle entering the atmosphere up to the
measurement of the EAS observable need to be modeled as precisely as
possible. Sophisticated EAS Monte Carlo simulation programs are
available for this task. This results in a highly indirect analysis
procedure.

In this article we will perform a detailed Monte Carlo study of
different methods to derive the proton-air cross section from air
shower data at an energy of $\unit[\sim10^{19}]{eV}$. We will assume
protons as cosmic ray particles in the energy range considered
here. Although there exist theoretical models~\cite{Berezinsky:2005cq,
  Berezinsky:2006nq,Berezinsky:2007wf, Zatsepin:1966jv,Greisen:1966jv}
and also experimental indications~\cite{Cronin:2007zz,Abbasi:2007sv,
  Abraham:2008ru} for this flux being indeed dominated by protons,
other elements in the primary flux have also to be taken into
account. This will be done in a forthcoming article that will address
specifically this topic~\cite{Ulrich:2009xxy}.

Based on the results of the Monte Carlo study of existing analysis
methods a general improvement is proposed and explicitly applied to
one of the methods. By accounting for the actually measured cross
section already in the simulation of showers, the systematic
uncertainty due to the limited knowledge of hadronic interactions is
significantly reduced.  The performance of the improved method is
thoroughly tested for the application to high quality data of the
depth of the shower maximum. Sources of systematic uncertainties of
the resulting cross section are discussed. It is shown that the
dependence on the hadronic interaction model, the most important
source of systematic uncertainties, can be significantly reduced by
incorporating the measured cross section in a consistent way in the
shower simulation.

The analyses of this work are done at an energy of
$\unit[10^{19}]{eV}$, at which high statistics data are available or
expected from HiRes~\cite{Abbasi:2004nz}, the Pierre Auger
Observatory~\cite{Abraham:2004dt}, and Telescope
Array~\cite{Kawai:2008zza}, but the results are qualitatively also
valid at other energies. The expected reduction of the model
dependence will be smaller at energies where data from accelerators
are available, i.e.\ $\unit[10^{15}]{eV}$ and below.

%%%%%%%%%%%%%%%%%%%%%%%%%%%%%%%%%%%%%%%%%%%%%%%%%%%%%%%%%%%%%%%%%%%%%%%%%%%%% 
\section{Relation between air shower observables and depth of the first interaction point}

%%%%%%%%%%%%%%%%%%%%%%%%%%%%%%%%%%%%%%%%%%%%%%%%%%%%%%%%%%%%%%%%%%%%%%%%%%%%%
\begin{figure}[t!]
  \centering
  \includegraphics[width=.75\linewidth]{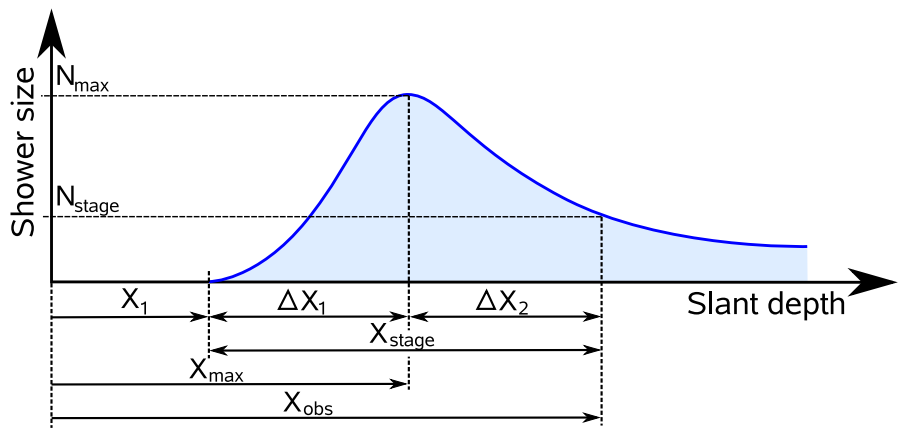}
  \caption{Definition of variables to characterize EAS longitudinal
    shower development.  Zero slant depth is where the cosmic ray
    particle enters the atmosphere.  The first interaction occurs at
    $X_1$. At $X_{\rm max}$ the shower reaches its maximum particle
    number $N_{\rm max}$. After the maximum the shower is attenuated
    over $\Delta X_2$ before it reaches the particle number $N_{\rm
      stage}$ at the slant depth of $X_{\rm obs}$.}
  \label{f:ShowerStages}
\end{figure}  
%%%%%%%%%%%%%%%%%%%%%%%%%%%%%%%%%%%%%%%%%%%%%%%%%%%%%%%%%%%%%%%%%%%%%%%%%%%%%

Interactions over many decades in energy are occurring during EAS
development. In the startup phase of the shower, relatively few
ultra-high energy hadronic interactions are distributing the energy of
the primary cosmic ray particle to a quickly growing number of
secondary particles. The stochastic nature of these initial
interactions is the source of strong fluctuations of EAS initiated by
identical cosmic ray particles (shower-to-shower fluctuations). A
significant part of secondary particles decay to electromagnetic
particles and lead to the development of an electromagnetic cascade
that, already after a few interactions, contains most of the total EAS
energy and constitutes by far the largest fraction of the
particles. The interactions of particles in the e.m.\ cascade are
theoretically well understood. Also the large number of these
interactions levels out additional large-scale
fluctuations\footnote{An exception are electromagnetic showers of $E >
  10^{18}$\,eV that, by chance, happen to start very deep in the
  atmosphere, for which the Landau-Pomeranchuk-Migdal effects leads to
  very large shower-to-shower fluctuations
  \cite{Landau:1953um,Landau:1953gr,Migdal:1956tc}.}.  Thus, the
development of an air shower can be characterized by two main stages:
\begin{itemize}
 \item[(i)]\textbf{Startup phase}, consisting of few initial hadronic
   interactions at ultra-high energies. These interaction are the main
   source of fluctuations.
 \item[(ii)]\textbf{Cascade phase}, in which very numerous
   interactions of the bulk of the shower particles at intermediate
   energies take place. No significant large-scale fluctuations are
   expected from this part.
\end{itemize}
A clear boundary between the two stages cannot be drawn. The
transition is seamless and by itself subject to strong
shower-to-shower fluctuations. Only treatment with full air shower
Monte Carlo simulation programs can fully account for all
fluctuations.

To discuss the relation of the depth of the first interaction point to
air shower observables it is useful to introduce a simple longitudinal
model of air shower development that connects the proton-air cross
section to the observables in a transparent way. The naming
conventions for the model are given in Fig.~\ref{f:ShowerStages}. We
distinguish between two types of air shower observation: by ground
based detector arrays and the direct observation of longitudinal
shower profiles by telescope detectors. The relevant air shower
observables used in proton-air cross section analyses are the position
of the shower maximum, $X_{\rm max}$, or the combination of the number
of electrons $N_{\rm e}$ and muons $N_{\mu}$ at the atmospheric slant
depth of the detector $X_{\rm obs}$.

In the following, the correlation of these EAS observables to the
characteristics of the ultra-high energy interactions is studied with
the one dimensional air shower Monte Carlo program
CONEX~v2r2~\cite{Bergmann:2006yz}.

The particle numbers are defined as the total number of electrons
above \unit[1]{MeV} and muons above \unit[1]{GeV}.  This corresponds
to typical quantities observed in air shower arrays, however the exact
definition of observables depends very much on the experimental setup
and varies strongly from experiment to experiment. By using the total
number of particles for our study we are focusing on general air
shower properties.

The shower maximum $X_{\rm max}$ is the slant depth of the maximum
energy deposit of the shower in the atmosphere. This definition
matches the shower maximum derived from the fluorescence light profile
of showers and coincides with that of the particle number within
$\sim\unit[3]{gcm^{-2}}$.

%%%%%%%%%%%%%%%%%%%%%%%%%%%%%%%%%%%%%%%%%%%%%%%%%%%%%%%%%%%%%%%%%%%%%%%%%%%%%
\subsection{Arrays of particle detectors}
%
%%%%%%%%%%%%%%%%%%%%%%%%%%%%%%%%%%%%%%
\begin{figure*}[t!]
  \centering
  \includegraphics[width=.615\linewidth]{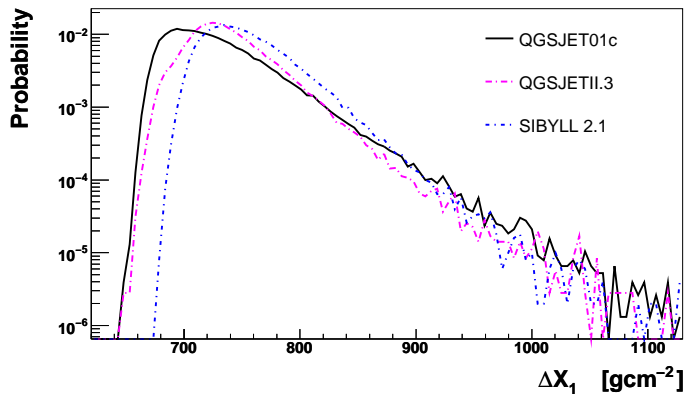}
  \vspace*{-.2cm}
  \caption{Distribution of the depth distance from the point of the
    first interaction up to the shower maximum for proton primaries at
    $\unit[10^{19}]{eV}$. The predictions of different models are
    compared.}
  \label{f:DeltaXmodel}
\end{figure*}
%%%%%%%%%%%%%%%%%%%%%%%%%%%%%%%%%%%%%%

Using ground based air shower arrays, one can estimate the proton-air
cross section by measuring the frequency of air showers of the same
energy and stage of their development at different atmospheric
depths. By selecting events of the same energy but different
directions of incidence, the point of the first interaction has to
vary with the angle, in order to observe the shower at the same stage
of development. The selection of showers of constant energy and stage
can be done only approximately and depends on the particular detector
setup, but the typical requirement is a constant $(N_{\rm e},
N_{\mu})$ at observation level.  Examples of measurements of this type
are~\cite{Hara:1983pa,Honda:1993kv,Aglietta:2009zz,Nikolaev:1993mc,Collaboration:2009ca}.
By requiring a given number of muons at the detector level does, in
first approximation, select EAS of the same primary energy, because
the attenuation of muons is small. However, also muons are slowly
attenuated in the atmosphere. To correct for that a constant intensity
selection can be applied~\cite{AlvarezMuniz:2002xs}.  Showers with
identical primary energy at the same stage of their shower development
are assumed to yield the same number of electrons since, after the
shower maximum, the electromagnetic shower attenuation is
approximately
universal~\cite{Rossi:1941zz,Billoir:2005,Nerling:2006yt,Giller:2005qz,%
  Lafebre:2009en,Lipari:2008td}.

The number of showers $N_{\rm sh}$ selected per area ${\rm d}A$ and time ${\rm d}t$, requiring a
constant $(N_e, N_\mu)$ at the atmospheric depth $X_{\rm obs}$ can be written as
\begin{eqnarray}
\frac{{\rm d}N_{\rm sh}}{{\rm d}N_{\rm e}\,{\rm d}N_\mu\,{\rm d}X_{\rm obs}\,{\rm d}A\,{\rm d}t} &=& 
  \int {\rm d}\Omega\, {\rm d}E\, {\rm d}X_1\, {\rm  d}\Delta X_1\, {\rm d}\Delta X_2\, 
\frac{{\rm e}^{-X_{\rm 1}/\lambda_{\rm int}}}{\lambda_{\rm int}}\,
\nonumber\\ 
  & &\times 
P_{\Delta X_1}(E)
~ P_{\Delta X_2}(E,X_1,\Delta X_1)
~ P_{N_\mu}(E,X_1,\Delta X_1,\dots)
\nonumber\\ 
 & &\times 
  \delta\bigl(X_{\rm obs}-X_1-\Delta X_1-\Delta X_2\bigr)
\nonumber\\ 
  & &\times 
  \frac{{\rm d}N_{\rm CR}}{{\rm d}\Omega\,{\rm d}E\,{\rm d}A\,{\rm d}t\,} \;,
  \label{eqn::frequency}
\end{eqnarray} 
where ${\rm d}N_{\rm CR}/({\rm d}\Omega\,{\rm d}E\,{\rm d}A\,{\rm
  d}t)$ is the flux of cosmic ray particles.  The distribution
$P_{\Delta X_1} = {\rm d}p_1/{\rm d}\Delta X_1$ describes 
shower-to-shower fluctuation of $\Delta X_1$, see
Fig.~\ref{f:DeltaXmodel}. The two-dimensional probability density
$P_{\Delta X_2}= {\rm d}p_2/({\rm d}\Delta X_2\,{\rm d}N_{\rm e})$ is
the frequency of showers of given energy $E$, $X_1$, and $\Delta X_1$ to
have a shower size $N_{\rm e}$ at a depth distance $\Delta X_2$ from
the shower maximum. It is required that the electron number after the
shower maximum is attenuated to $N_{\rm e}$, while $X_{\rm
  obs}=X_1+\Delta X_1+\Delta X_2$.  The selection of showers by energy
according to their muon number is reflected by the energy dependent
probability distribution $P_{N_\mu}= {\rm d}p_\mu/{\rm d}N_\mu $.  

%%%%%%%%%%%%%%%%%%%%%%%%%%%%%%%%%%%%%%
\begin{figure*}[t!]
  \centering
  \includegraphics[width=.615\linewidth]{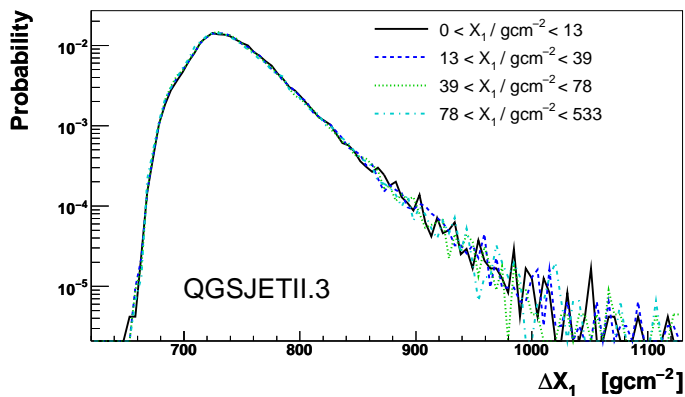}~
  \vspace*{-.2cm}
  \caption{Independence of $P_{\Delta X_1}$ of $X_{\rm 1}$,
    demonstrated for primary protons at $\unit[10^{19}]{eV}$ with
    \textsc{QGSJetII}.}
  \label{f:DeltaXuniversality}
\end{figure*}
%%%%%%%%%%%%%%%%%%%%%%%%%%%%%%%%%%%%%%

It is important to notice that the distribution $P_{\Delta X_1}$ does
not depend on the depth $X_1$ of the first interaction point. This is
shown in Fig.~\ref{f:DeltaXuniversality} for one hadronic interaction
model. At high energy, and for air densities of relevance, almost all
charged secondary pions always interact and all neutral pions decay
immediately. Furthermore, the electromagnetic cascade initiated by the
photons from $\pi^0$ decay does not depend on the local air
density. This makes the startup phase of a shower to a good
approximation independent of $X_1$.

\begin{figure*}[t!]
  \centering
  \includegraphics[width=.33\linewidth]{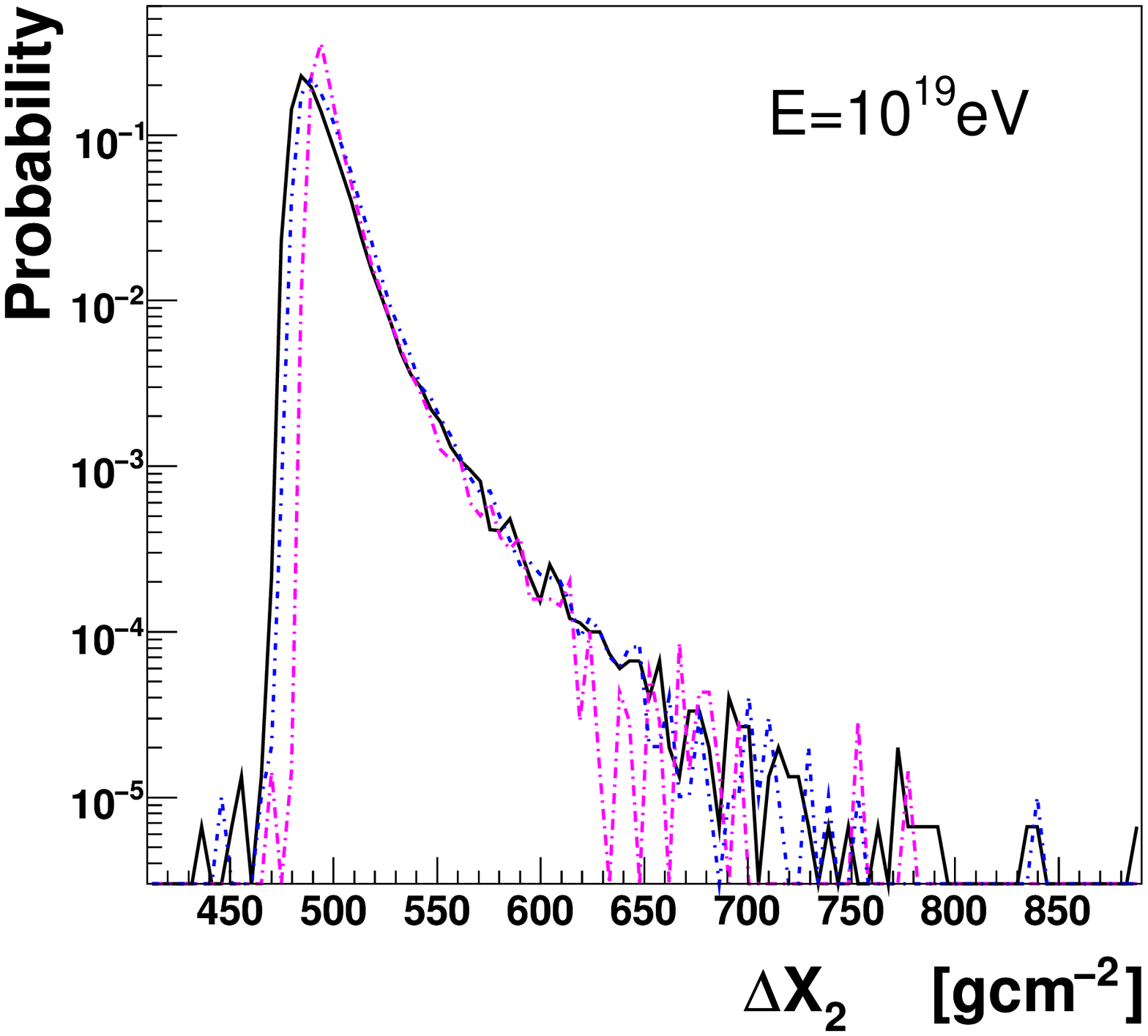}~
  \includegraphics[width=.33\linewidth]{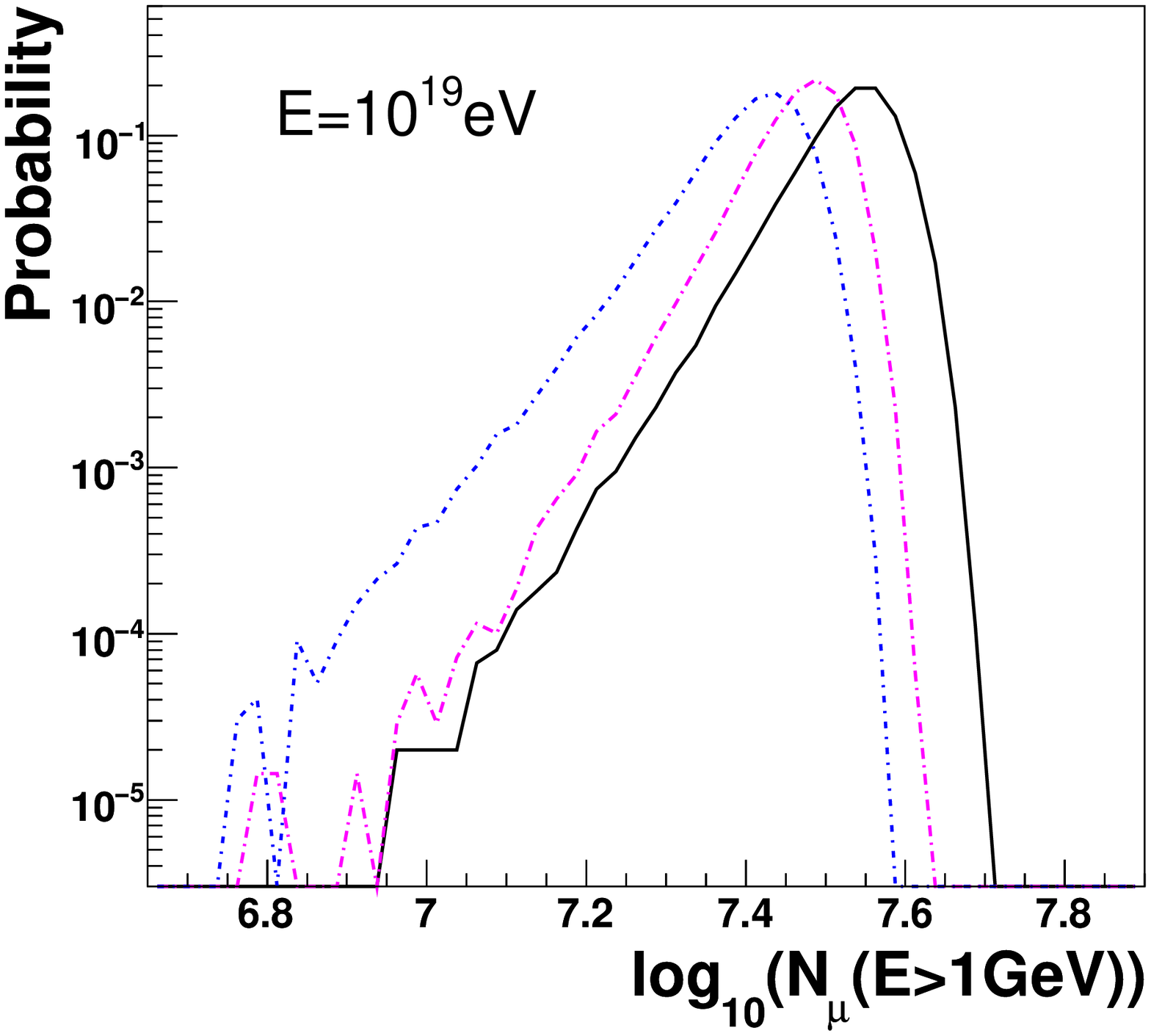}~
  \includegraphics[width=.33\linewidth]{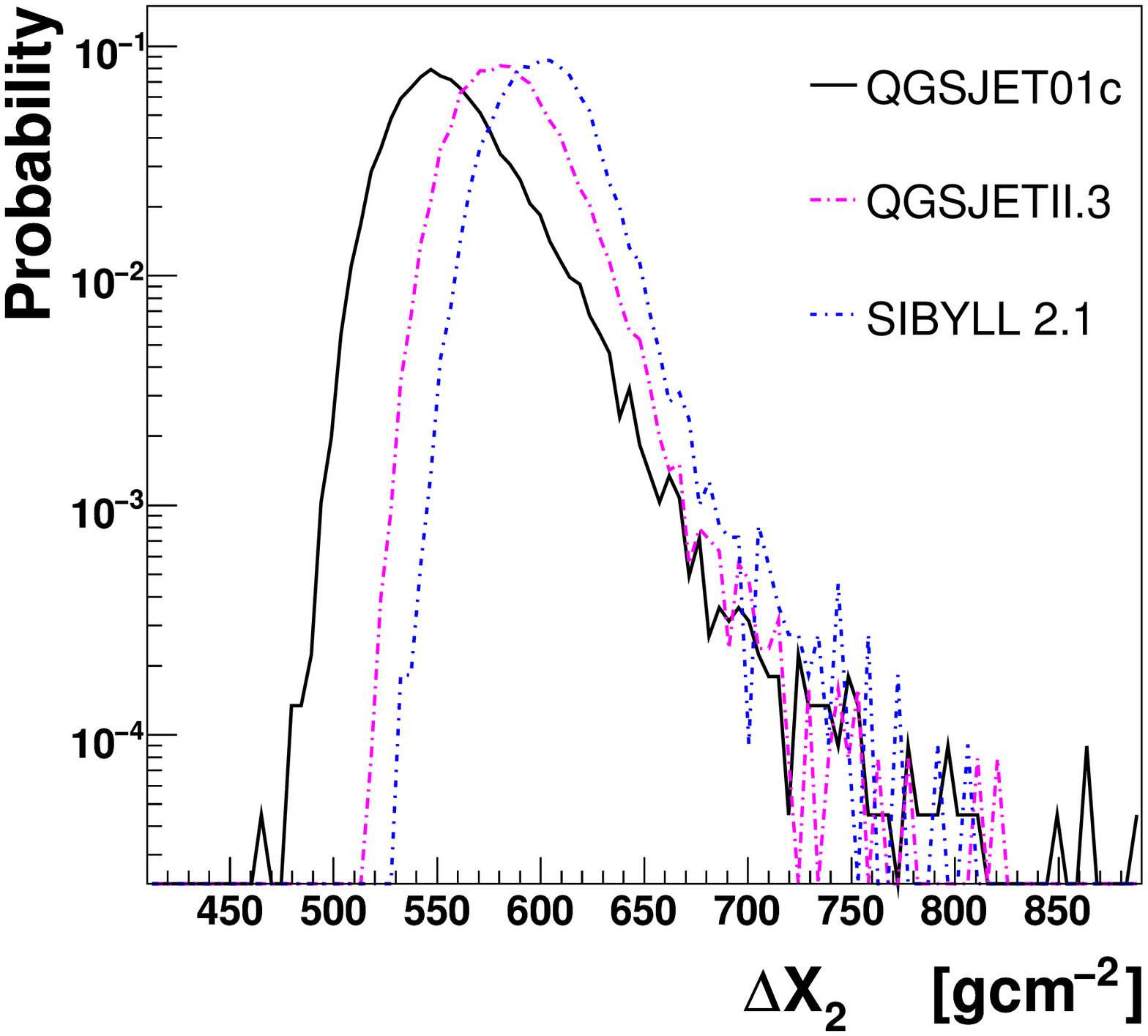}~
  \vspace*{-.2cm}
  \caption{Model dependence of longitudinal air shower development for
    proton primaries. Left panel: Slant depth distance from the shower
    maximum up to the depth where showers are attenuated to $N_{\rm
      e}(E_{\rm e}>\unit[1]{MeV})=10^9$ electrons.  Middle panel:
    Number of muons above \unit[1]{GeV} after $\unit[1000]{gcm^{-2}}$
    of shower development. Right panel: Same as left panel, but for
    showers selected according to
    $7.6<\log_{10}N_\mu(E_\mu>\unit[1]{GeV})<7.7$. While the plots on
    the left and in the middle are obtained from fixed energy
    simulations, the right plot is based on an energy spectrum
    $\propto E^{-3}$.}
  \label{f:DeltaXModelsS}
\end{figure*}

The distribution $P_{\Delta X_1}$ depends, however, on the hadronic
interaction model used for simulation. This is displayed in
Fig.~\ref{f:DeltaXmodel}: both the mean $\Delta X_1$ and the shape of
the distribution depend strongly on the chosen interaction model.  In
contrast, the distribution $P_{\Delta X_2}$ exhibits only a small
model dependence due to the universality of electromagnetic air
showers after shower maximum, see Fig.~\ref{f:DeltaXModelsS} (left).

An additional source of strong model dependence of
Eq.~(\ref{eqn::frequency}) is the number of muons predicted for a
given shower energy and depth. This can be seen in
Fig.~\ref{f:DeltaXModelsS}~(middle) where the frequency of finding a
certain muon number for showers at $10^{19}$\,eV and different models
is shown. The effect of this model dependence is displayed in
Fig.~\ref{f:DeltaXModelsS}~(right) by showing the folded distributions
\begin{equation}
\left. P_{\rm eff}\right|_{N_e,N_\mu} \propto \int  P_{\Delta X_2}~ P_{N_\mu}~ E^{-3}~{\rm d}E \,.
\end{equation}

In real data analysis it must be taken into account that $X_{\rm
  obs}$, $N_{\rm e}$ and $N_\mu$ are only known with a limited
precision due to the detector and shower reconstruction resolution. In
the model described by Eq.~(\ref{eqn::frequency}) this would
furthermore add the corresponding uncertainty distributions $P_{\rm
  res}^{\rm S}$ and integrations over the three observables. The
influence of the measurement resolution will be discussed below but is
not included in (\ref{eqn::frequency}) for sake of clarity.

% In existing analyses Eq.~(\ref{eqn::frequency}) was always formulated
% according to the zenith angle
% \begin{equation}
%   \frac{{\rm d}p}{{\rm d}N_{\rm e}\,{\rm d}N_\mu\,{\rm d}\sec\theta}
%  \simeq X_{\rm obs}^{\rm vert}\,\frac{{\rm d}p}{{\rm d}N_{\rm e}\,{\rm d}N_\mu\,{\rm
%       d}X_{\rm obs}}\;,
% \end{equation}
% which is only approximately applicable for zenith angles smaller than
% \unit[60]{$^\circ$}. Correctly, the curvature of the atmosphere must
% be considered.

%%%%%%%%%%%%%%%%%%%%%%%%%%%%%%%%%%%%%%%%%%%%%%%%%%%%%%%%%%%%%%%%%%%%%%%%%%%%%
\subsection{Optical telescopes}
\label{sec:XmaxObs}

Using fluorescence telescopes, the distribution of $X_{\rm max}$ can
be measured directly to obtain a handle on the value of the cross
section at the highest
energies~\cite{Ellsworth:1982kv,Baltrusaitis:1985mx}.

The direct observation of the position of the shower maximum allows us
to simplify Eq.~(\ref{eqn::frequency}) by removing the term $P_{\Delta
  X_2}$ describing the shower development after the shower maximum and
the distribution $P_\mu$ has to be replaced by a corresponding energy
estimator based on the longitudinal shower profile.  The resulting
distribution of $X_{\rm max}$ can be written as
\begin{eqnarray}
  \label{eqn::XmaxObs}
  \frac{{\rm d}N_{\rm sh}}{{\rm d}X_{\rm max} {\rm d}E_{\rm em} {\rm d}A {\rm d}t} &=& 
  \int {\rm d}\,\Omega\, {\rm d}E \, {\rm d}X_{\rm 1} \, {\rm d}\Delta X_{\rm 1} 
\; \frac{{\rm e}^{-X_{\rm 1}/\lambda_{\rm
        int}}}{\lambda_{\rm int}}\, 
P_{\Delta X_1}(E)\;
P_{E_{\rm em}}\,
\nonumber\\
  & &~~\times \delta\bigl(X_1+\Delta X_1 - X_{\rm
    max}\bigr) \,\frac{{\rm
      d}N_{\rm CR}}{{\rm d}A\, {\rm d}t\, {\rm d}\Omega\,{\rm d}E }\;.
\label{eqn::fdfrequency}
\end{eqnarray} 
Here $E_{\rm em}$ denotes the electromagnetic, i.e.\ calorimetric
energy of the shower that can be obtained from the integration of the
energy deposit profile.  There is a very good correlation between the
primary energy and the calorimetric energy \cite{Barbosa:2003dc}. This
correlation and the weak energy dependence of the depth of shower
maximum allow us to neglect $P_{E_{\rm em}} = {\rm d}p_{\rm em}/{\rm
  d}E_{\rm em}$ in our numerical studies.

\section{Analysis of cross section measurement methods}
\label{sec:ProtonAirCX}

The aim of all methods to measure the cross section is the
determination of the interaction length $\lambda_{\rm int}$ from
measured distributions that represent the l.h.s. of
Eqs.(\ref{eqn::frequency},\ref{eqn::fdfrequency}).

%%%%%%%%%%%%%%%%%%%%%%%%%%%%%%%%%%%%%%%%%%%%%%%%%%%%%%%%%%%%%%%%%%%%%%%%%%%%% 
\subsection{``$k$-factor'' techniques}
The approximation of an exponential attenuation of the frequency of
air showers after the penetration of large amounts of atmosphere is
the basis of the $k$-factor
method~\cite{Hara:1983pa,Ellsworth:1982kv,Baltrusaitis:1985mx}, which
has been used to analyze both $X_{\rm max}$ and \NeNmu~data.  The
exponential slope of the attenuation is typically denoted by
$\Lambda$, which is then related to the hadronic interaction
length by a so-called $k$-factor
\begin{equation}
  \Lambda = k\,\lambda_{\rm p-air} \;.
\end{equation}
While this method has been applied to data of air shower arrays as
well as telescope detectors~\cite{Honda:1993kv,%
  Nikolaev:1993mc,Aglietta:2009zz,Hara:1983pa,Collaboration:2009ca,%
  Knurenko:1999cr,Baltrusaitis:1984ka}, the definition of the
$k$-factor for a ground array, $k_{\rm S}$, and for a telescope
detector, $k_{\rm X}$, are not identical. Air shower fluctuations
enter differently into $k_{\rm S}$ and $k_{\rm X}$ and the detector
resolutions are also very different. This can be understood based on
Eq.~(\ref{eqn::frequency}) and (\ref{eqn::fdfrequency}), which can
both be approximated by exponential distributions for large $X_{\rm
  max}$, respectively $X_{\rm obs}$. For the $(N_{\rm e},N_\mu)$
method this corresponds to
\begin{equation}
  \label{eqn:LambdaSD}
  \left.\frac{{\rm d}N_{\rm sh}}{{\rm d}\sec\theta}\right|_{N_{\rm e}, N_{\mu}} \;\propto\; {\rm e}^{-X_{\rm
      obs}/\Lambda_{\rm S}} \;\propto\; {\rm
    e}^{-X_{\rm obs}^{\rm vert}\sec\theta/\Lambda_{\rm S}} \;, 
\end{equation}
while for the $X_{\rm max}$-tail method it is
\begin{equation}
  \label{eqn:telescopeAtt}
  \frac{{\rm d}N_{\rm sh}}{{\rm d}X_{\rm max}}\;\propto\; {\rm
    e}^{-X_{\rm max}/\Lambda_{\rm X}} \;.
\end{equation}
This assumes that there are no significant non-exponential
contributions to the tails of the distributions, which is not true in
general as it will be shown in the following.  Since the full
distributions are described by Eq.~(\ref{eqn::frequency}) and
(\ref{eqn::fdfrequency}) it is possible to calculate the tails
based on the given convolution integrals. Each of the integrations
will give a contribution, so for $(N_{\rm e},N_\mu)$ one gets 
\begin{equation} 
  \label{eqn:FrequencyLambda}
  \Lambda_{\rm S} = \lambda_{\rm int} \, k_{\Delta X_{\rm 1}}
  \, k_{\Delta X_{\rm 2}} \, k_{\rm res}^{\rm S} = \lambda_{\rm
    int} \, k_{\rm S} 
  \qquad \mbox{with} \qquad k_{\rm S}=k_{\Delta X_{\rm 1}}
  \, k_{\Delta X_{\rm 2}} \, k_{\rm res}^{\rm S} \;,
\end{equation}      
while for $X_{\rm max}$-tail just
\begin{equation}
  \label{eqn:XmaxObsLambda}
  \Lambda_{\rm obs}^{\rm X} = \lambda_{\rm int} \, k_{\Delta X_{\rm
      1}} \, k_{\rm res}^{\rm X} = \lambda_{\rm int} \, k_{\rm X}
  \qquad\qquad \mbox{with} \qquad k_{\rm X}=k_{\Delta X_{\rm 1}} \,
  k_{\rm res}^{\rm X} \,,
\end{equation}
where $\lambda_{\rm int}$ is the proton-air interaction length,
$k_{\Delta X_1}$ the contribution from the integration of $P_{\Delta
  X_1}$, $k_{\Delta X_2}$ the one from $P_{\Delta X_2}$ and $k_{\rm
  res}^{\rm S/X}$ the part due to the detector resolution.  The
individual contributions of $k_{\Delta X_1}$, $k_{\Delta X_2}$ and
$k_{\rm res}^{\rm S/X}$ are difficult to compute and not known in most
of the existing analyses, except the one in
Ref.~\cite{Knurenko:1999cr} and even more complete in
  Ref.~\cite{Collaboration:2009ca}.

%%%%%%%%%%%%%%%%%%%%%%%%%%%
\begin{figure*}[tb]
  \includegraphics[width=.48\textwidth]{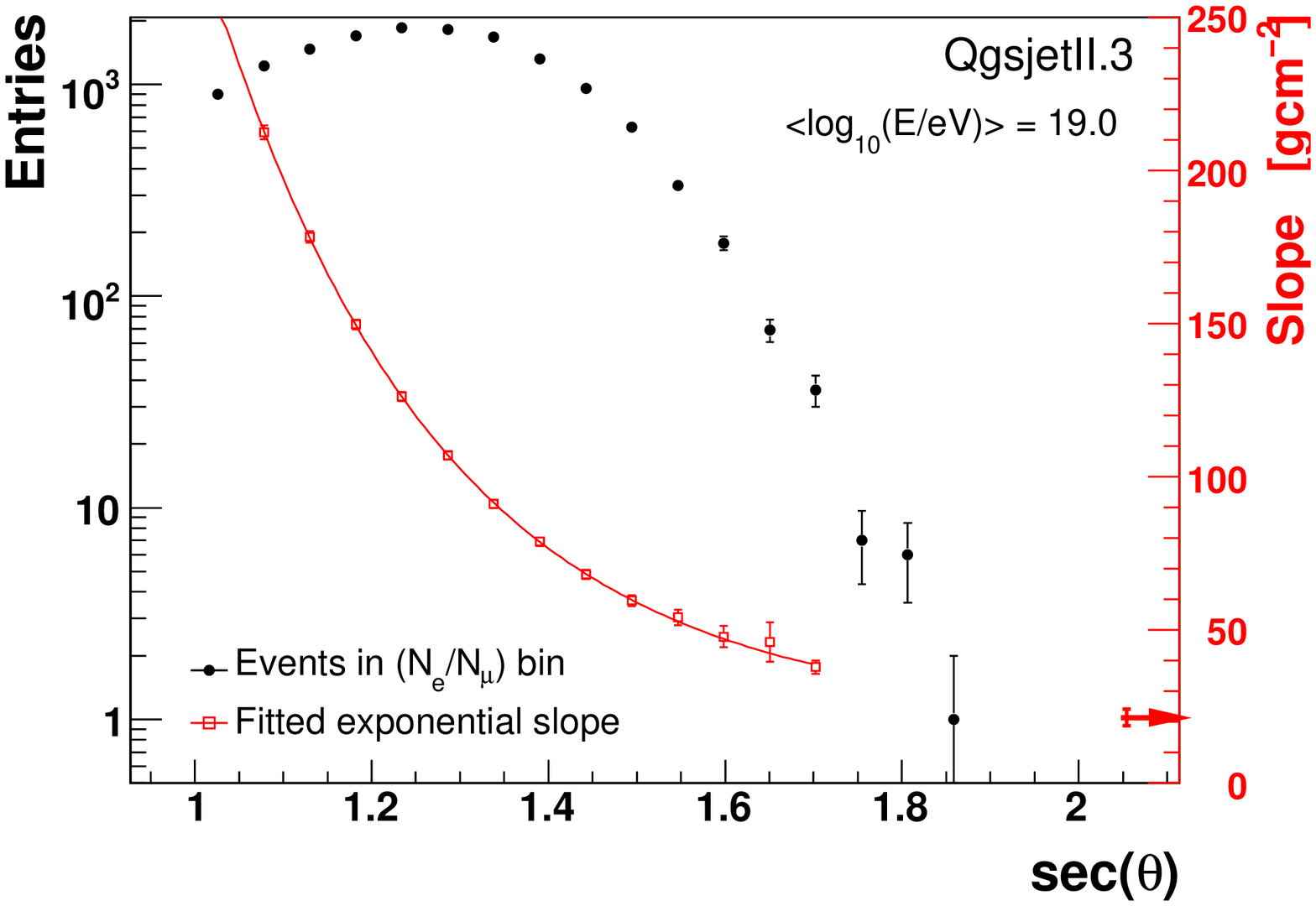}%
  \hfill%
  \includegraphics[width=.48\textwidth]{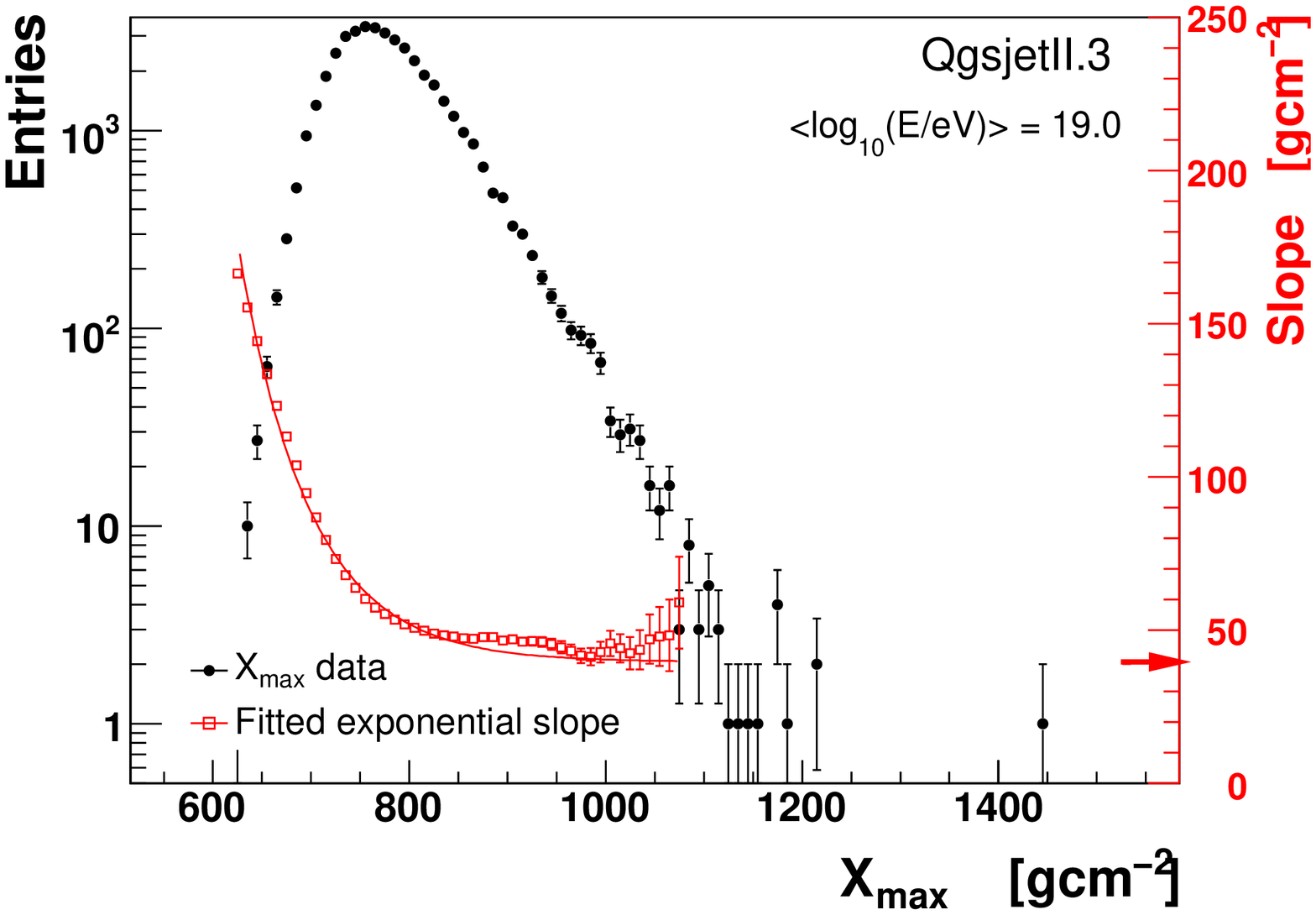}\\
  \vspace*{-.4cm}
  \caption{Left panel: $(N_{\rm e},N_\mu)$-method. Right panel: $X_{\rm
      max}$-tail method. The fitted open symbols show the dependence of the
    resulting exponential slope on the choice of the begin of the fit
    range. The arrow indicates the asymptotic behavior of the fit,
    deduced from Eq.~(\ref{eq:lambda_fit}), and its statistical
    uncertainty.}
  \label{fig:tail_vs_frequency}
\end{figure*}
%%%%%%%%%%%%%%%%%%%%%%%%%%%

%
For a given experimental setup and analysis approach it is possible to
estimate the values of $k_{\Delta X_1}$, $k_{\Delta X_2}$, $k_{\rm
  res}^{\rm X}$ and $k_{\rm res}^{\rm S}$ as well as their dependence
on hadronic interaction models. This is demonstrated here for an
artificial experimental setup as described below. For each of the hadronic interaction
models \textsc{QGSJet01c}, \textsc{QGSJetII.3} and \textsc{SIBYLL 2.1}
a set of $\sim400,000$ proton induced air
showers is simulated with a primary energy distribution
${\propto}E^{-3}$ between $18.7<\log_{10}E/\rm eV<19.4$ and
a zenith angle distribution of ${\rm d}N\propto\cos\theta\,{\rm
  d}\cos\theta$.  Two histograms are produced and analyzed for every
set of simulations:
\begin{itemize}
\item Distribution of $X_{\rm max}$ with $18.9<\log_{10}E/\rm
  eV<19.15$. A Gaussian detector resolution with $\sigma(X_{\rm
    max})=\unit[20]{gcm^{-2}}$ and an energy resolution of
  $\sigma(E)/E=0.1$ is assumed, which corresponds to values reported
  by the Pierre Auger Collaboration~\cite{Dawson:2007di}. The
  resulting mean energy of this data sample is $\unit[10^{19.0}]{eV}$.
\item Distribution of events with $1.26<N_{\rm e}/10^9<3.98$ and
  $0.28<N_\mu/10^8<1.12$ at a vertical atmospheric depth of $X_{\rm
  obs}^{\rm vert}=\unit[860]{gcm^{-2}}$ (corresponding to the Pierre
  Auger Observatory) versus $\sec\theta$.  The $N_{\rm e}$ cut selects
  showers past their maximum ($N_{\rm e}^{\rm
    max}(\unit[10^{19}]{eV})\sim6.3\cdot10^9$) as done, for example,
  in the pioneering Akeno analysis~\cite{Hara:1983pa}\footnote{In the
    recent analysis of the EAS-TOP Collaboration, showers are selected
    close to their maximum development~\cite{Aglietta:2009zz}.}.  The
  detection resolution of $N_{\rm e}$ and $N_\mu$ are assumed to be
  Log-Normal with a resolution of $\sigma(\log_{10}N_{\rm
    e})/\log_{10}N_{\rm e}=0.05$ and
  $\sigma(\log_{10}N_{\mu})/\log_{10}N_{\mu}=0.1$ as well as the
  zenith angle uncertainty of $\sigma(\theta)=1.0^\circ$. Due to
  significant model-differences in the prediction of muon numbers the
  selected data samples have slightly differing energies (from
  $\unit[10^{18.96}]{eV}$ for \textsc{QGSJet01} up to
  $\unit[10^{19.04}]{eV}$ for \textsc{SIBYLL}).
\end{itemize}
The histograms are fitted with an exponential function by a
log-likelihood fit, which allows to correctly include empty bins in
the analysis.  The fit range is chosen as follows: While the end of
the fit, $X^{\rm fit}_{\rm end}$, is always $\unit[50]{gcm^{-2}}$
beyond the last non-zero entry in the histogram, the start of the fit,
$X^{\rm fit}_{\rm start}$, is varied. The resulting dependence of the
exponential slopes on the start of the fitting range is then
parameterized by
\begin{equation} 
  \label{eq:lambda_fit} 
  \Lambda(X^{\rm fit}_{\rm start}) = \Lambda_0 + \Lambda_1\,{\rm e}^{-c\,X^{\rm
      fit}_{\rm start}}\;,
\end{equation} 
where the asymptotic slope $\Lambda_0$ for $X^{\rm fit}_{\rm
  start}\rightarrow\infty$ is taken to be the fit range-independent
value of the exponential slope. In Fig.~\ref{fig:tail_vs_frequency}
this procedure is displayed for histograms obtained with the
\textsc{QGSJetII} interaction model. It is found that the choice of
the fit range has a major impact on the outcome of the slope. This is
consistent with the results of~\cite{AlvarezMuniz:2002xs}.  For the
($N_{\rm e}$,$N_\mu$)-method, generally no plateau is found and the
resulting value for $\Lambda_0$ is highly unstable. Thus, it is hardly
possible to define a meaningful slope of the tail for this case. The
situation for $X_{\rm max}$ is somewhat better. While a stable plateau
is still difficult to identify, $\Lambda_0$ can be estimated
reliably. With this analysis approach the resulting asymptotic values,
$\Lambda_0$, are slightly lower than the values obtained for $\Lambda$
directly from fits to the distributions.  So neither the slopes nor
the $k$-factors can be directly compared to results from previous
analyses. Nevertheless, the found model dependence and other
methodical problems will apply in a similar way to previous analyses.
The strong dependence of the slope on the chosen fitting range is a
severe methodical problem of the $k$-factor technique. Without a given
method to infer a meaningful slope from such distributions, as for
example the one proposed here, the $k$-factor technique cannot produce
reliable results. 

The obtained slopes $\Lambda_0$ are then compared to the interaction mean free
path length of the primary particle in the atmosphere, $\lambda_{\rm
  p-air}$, of the interaction model used in the CONEX simulations.  In
Tab.~\ref{tab:tail_vs_frequency} the resulting $k$-factors are listed
together with their propagated statistical uncertainties. The total
model induced uncertainties on the $k_{\rm S/X}$-factors are
$\unit[\sim7]{\%}$ for the $X_{\rm max}$-tail and $\unit[\sim28]{\%}$
for the \NeNmu-method.  Looking to the $k$-factor components one can
see that a part of the model induced uncertainty enters already in the
shower development up to the shower maximum ($k_{\Delta X_1}$), while
the shower development after the shower maximum ($k_{\Delta X_2}$)
adds the more significant amount of model-dependence. Another large
contribution comes from $k_{\rm res}^{\rm S}$, which is mostly related
to very differing model predictions on the number of muons.  The
factor $k_{\rm res}^{\rm X}$ is only marginally model-dependent, which
is originating from the very small model differences in the energy
dependence of $X_{\rm max}$.

While this study mostly serves illustrative purposes, and the chosen
experimental and analysis parameters are arbitrary, it nevertheless
demonstrates the impact of model dependence on $k$-factor techniques. The
found model dependence is rooted in the underlying air shower physics as well
as in typical detector characteristics.
\begin{table*}[bt!]
  \caption{Resulting $k$-factors from CONEX study for primary protons at
    $\unit[10^{19}]{eV}$. In each column the maximal and minimal $k$-factors are
    highlighted. The difference between those maxima is denoted by $\Delta$
    and is a measure of model-dependence.}
  \vspace*{.2cm}
  \footnotesize
 \begin{tabular}{r||cccc|cc}
 Model & $k_{\Delta X_1}$ & $k_{\Delta X_2}$ & $k_{\rm res}^{\rm X}$ & $k_{\rm res}^{\rm S}$ & $k_{\rm X}$ & $k_{\rm S}$ \\
  \hline\hline
\textsc{QgsjetII.3} & \textbf{1.00$\pm$0.01} & {0.61$\pm$0.06} & \textbf{0.92$\pm$0.01} & \textbf{0.81$\pm$0.10} & \textbf{0.92$\pm$0.01} & \textbf{0.49$\pm$0.06} \\
\textsc{Qgsjet01c} & \textbf{1.14$\pm$0.00} & \textbf{0.48$\pm$0.05} & \textbf{0.93$\pm$0.01} & \textbf{0.50$\pm$0.14} & \textbf{1.06$\pm$0.01} & \textbf{0.27$\pm$0.08} \\
\textsc{Sibyll 2.1} & {1.02$\pm$0.00} & \textbf{0.68$\pm$0.06} & {0.93$\pm$0.01} & {0.61$\pm$0.11} & {0.94$\pm$0.01} & {0.42$\pm$0.08} \\
\hline
Mean & 1.05$\pm$0.00 & 0.59$\pm$0.03 & 0.92$\pm$0.00 & 0.64$\pm$0.07 & 0.97$\pm$0.00 & 0.40$\pm$0.04 \\
$\Delta$ & 0.07$\pm$0.00 & 0.10$\pm$0.04 & 0.01$\pm$0.00 & 0.16$\pm$0.09 & 0.07$\pm$0.00 & 0.11$\pm$0.05 \\
\end{tabular}
  \label{tab:tail_vs_frequency}
\end{table*}
%

%%%%%%%%%%%%%%%%%%%%%%%%%%%%%%%%%%%%%%%%%%%%%%%%%%%%%%%%%%%%%%%%%%%%%%%%%%%%%
\subsection{Unfolding of the $X_{\rm max}$-distribution}
\label{sec:XmaxUnfolding}
An improvement of the cross section measurement techniques is achieved
by explicitly accounting for air shower
fluctuations~\cite{Belov:2006mb}. This allows one to use not only the
slope but also the shape of the $X_{\rm max}$-distribution. The
measured $X_{\rm max}$-distribution, Eq.~(\ref{eqn::XmaxObs}), is
unfolded using a given $P_{\Delta X_{\rm 1}}$-distribution to retrieve
the original $X_{\rm 1}$-distribution.  Of course, the $\Delta
X_1$-distribution needs to be inferred from simulations, which
ultimately introduces a comparable model dependence as in the
$k$-factor techniques~\cite{Ulrich:2006hb}.  The model dependence of
the most important part of the kernel function, $P_{\Delta X_1}$, can
be seen in Fig.~\ref{f:DeltaXmodel}.

In the unfolding technique, a larger range of the $X_{\rm
  max}$-distribution is used. If the primary particles are known to be
protons only, a cross section analysis can be done already with very limited
shower statistics. On the other hand, the results of this method are
more sensitive to a possible fraction of primary particles that are
not protons.

%%%%%%%%%%%%%%%%%%%%%%%%%%%%%%%%%%%%%%%%%%%%%%%%%%%%%%%%%%%%%%%%%%%%%%%%%%%%%
%%%%%%%%%%%%%%%%%%%%%%%%%%%%%%%%%%%%%%%%%%%%%%%%%%%%%%%%%%%%%%%%%%%%%%%%%%%%%
%%%%%%%%%%%%%%%%%%%%%%%%%%%%%%%%%%%%%%%%%%%%%%%%%%%%%%%%%%%%%%%%%%%%%%%%%%%%%
\section{An improved method to derive the proton-air cross section}
\label{sec:method}

One of the shortcomings of cross section analysis methods applied so
far is the missing connection between the cross section of the first
interaction, that is to be measured, and the cross sections used in
the calculation of the various probability distributions in
Eqs.~(\ref{eqn::frequency}) and (\ref{eqn::fdfrequency}).

In the following we consider the method of unfolding the $X_{\rm max}$
distribution, applicable to fluorescence telescope measurements, and
improve it by consistently accounting for the ``measured'' high-energy
cross section.  We will calculate the full shape of the distribution
of $X_{\rm max}$, depending only on properties of hadronic
interactions at ultra-high energies.  The impact of changing features
of hadronic interactions at ultra-high energies on the longitudinal
air shower development, i.e.\ $P_{\Delta X_1}$, is parameterized and
used within the calculation.  It is straightforward to refine the
modeling by incorporating detector acceptance as well as the energy
distribution of analyzed data.

\subsection{Description of the $X_{\rm max}$-distribution}

The description of the distribution of observed $X_{\rm
  max}^{\rm rec}$ in terms of $\sigma_{\rm p-air}$ is based on
Eq.~(\ref{eqn::XmaxObs}), now written with the term for the experimental $X_{\rm max}$ resolution 
\begin{eqnarray}
  \frac{{\rm d}N_{\rm sh}}{{\rm d}X_{\rm max}^{\rm rec}} \propto \int {\rm d}X_{\rm 1}
       \,{\rm d}\Delta X_{\rm 1}\,
&&\, {\rm d}X_{\rm max} \;\; \frac{{\rm e}^{-X_{\rm
        1}/\lambda_{\rm int}}}{\lambda_{\rm int}}\; 
P_{\Delta  X_1}(E,\lambda_{\rm int},\Delta X_1+X_{\rm shift})
\nonumber\\
  && \times\,\delta(X_1+\Delta X_1-X_{\rm max}) \; P_{\rm res}^{\rm X}\bigl(X_{\rm max}^{\rm
    rec}\,|\,X_{\rm max}\bigr) \;.
\label{eqn::improved}
\end{eqnarray} 
The parameter $X_{\rm shift}$ allows us to shift the $P_{\Delta
  X_1}$-distribution by replacing $\Delta X_{\rm 1}$ with $\Delta
X_{\rm 1}+X_{\rm shift}$.  The introduction of the $X_{\rm shift}$
parameter is necessary in order to reduce the model dependence of the
analysis~\cite{Ulrich:2006hb}.  This can be easily understood by
looking at the $P_{\Delta X_1}$-distributions shown in
Fig.~\ref{f:DeltaXmodel}, which appear to be shifted between different
models by up to $\unit[\sim60]{gcm^{-2}}$.  It is found that the
$X_{\rm shift}$ parameter is highly model dependent and comprises many
additional effects of the characteristics of high energy hadronic
interactions on the $X_{\rm max}$-distribution. It can be demonstrated
that differences between the inelasticity or the secondary
multiplicity of the models may act as a cause for such an additional
shift~\cite{Ulrich:2009xxx}. The only important assumption related to
the role of $X_{\rm shift}$ in the cross section analysis is that any
additional and unknown changing characteristics of hadronic
interactions at extreme energies contributes mainly to a global shift
of $X_{\rm max}$ and thus $\Delta X_{\rm 1}$, while leaving the shape
of the distributions unchanged.

%%%%%%%%%%%%%%%%%%%%%%%%%%%%%
\begin{figure}[tb!]
  \centering
  \includegraphics[width=.65\textwidth]{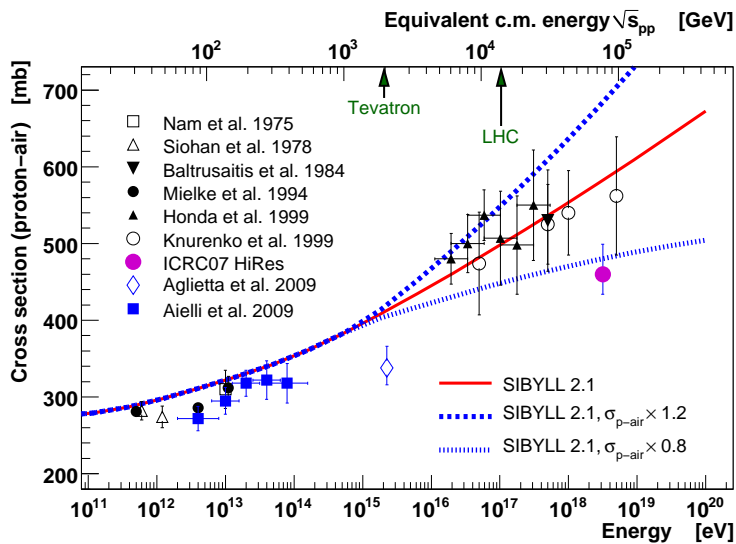}
  \caption{Modified extrapolation of the proton-air cross
    section. Shown are the original prediction from \textsc{SIBYLL}
    together with a $\unit[20]{\%}$ increased respectively decreased
    cross section at $\unit[10^{19}]{eV}$.}
  \label{fig:SigmaModifiedCrossSection}
\end{figure}
%%%%%%%%%%%%%%%%%%%%%%%%%%%%%

The cross section analysis proceeds then as follows. The $X^{\rm
  rec}_{\rm max}$-distribution is calculated from
Eq.~(\ref{eqn::improved}) for a given interaction length $\lambda_{\rm
  int}$ and shift parameter $X_{\rm shift}$ and compared to the
measured distribution. By performing a log-likelihood fit with the
interaction length and an overall shift in depth as free parameters,
the cross section and the $1\sigma$ uncertainty band is
found~\cite{Ulrich:2006hb}.

One major improvement with respect to previous cross section analysis
approaches is the consideration of the impact of a changing cross
section on the resulting shower development described by $P_{\Delta
  X_1}$. It is assumed that the cross section is reasonably well known
at the energy corresponding to that of the Tevatron collider. Starting
from this energy of $\sim 2\times 10^{15}$\,eV, cross sections of the
model used for the calculation of $P_{\Delta X_1}$ are rescaled by a
factor $f(E)$ that increases logarithmically with energy and ensures
that the modified model cross section matches the corresponding value
of the fit parameter $\lambda_{\rm int}$ at the considered shower
energy.  Here we will use $10^{19}$\,eV as reference for the
scaling parameter being $f_{19}$ at this energy. Then the scaling factor reads
\begin{equation}
f(E) = 1 + (f_{19} -1) \frac{\ln (E/10^{15}\,{\rm eV})}{\ln (10^{19}\,{\rm eV}/10^{15}\,{\rm eV})},
\label{eq:rescalingFactor}
\end{equation}
for $E> 10^{15}$\,eV and $f(E) = 1$ otherwise.  The idea of the
rescaling of the cross section is shown in
Fig.~\ref{fig:SigmaModifiedCrossSection}, where the proton-air cross
section of \textsc{SIBYLL} is shown together with cross sections
that are scaled up and down by 20\,\% at $10^{19}$\,eV.

So far the dependence of $P_{\Delta X_1}$ on the cross section has
been neglected in air shower based cross section measurement. In fact,
any attempt to determine the proton-air cross section without taking
this dependence into account is overestimating the impact of
$\sigma_{\rm p-air}$ on the analyzed observables since part of the
measured effect must be attributed to a modified development of the
air shower and not to the fluctuations of the first interaction point.

To implement the idea of modified cross sections in the data analysis
one needs to parameterize the $\Delta X_1$-distribution and determine
its energy dependence and the modification of this distribution as
function of the cross section scaling parameter. This is done using a
Moyal distribution extended by one parameter and described in detail
in~\ref{appendix}.  In Fig.~\ref{f:moyalDeltaX} example $\Delta
X_1$-distributions are shown together with fits of the extended Moyal
distribution and also the final parameterizations.

%%%%%%%%%%%%%%%%%%%%%%%%%%%%%
\begin{figure*}[t!]
  \centering
  \hfill
 \subfigure[\textsc{QGSJetII.3}]{\includegraphics[width=.45\linewidth]{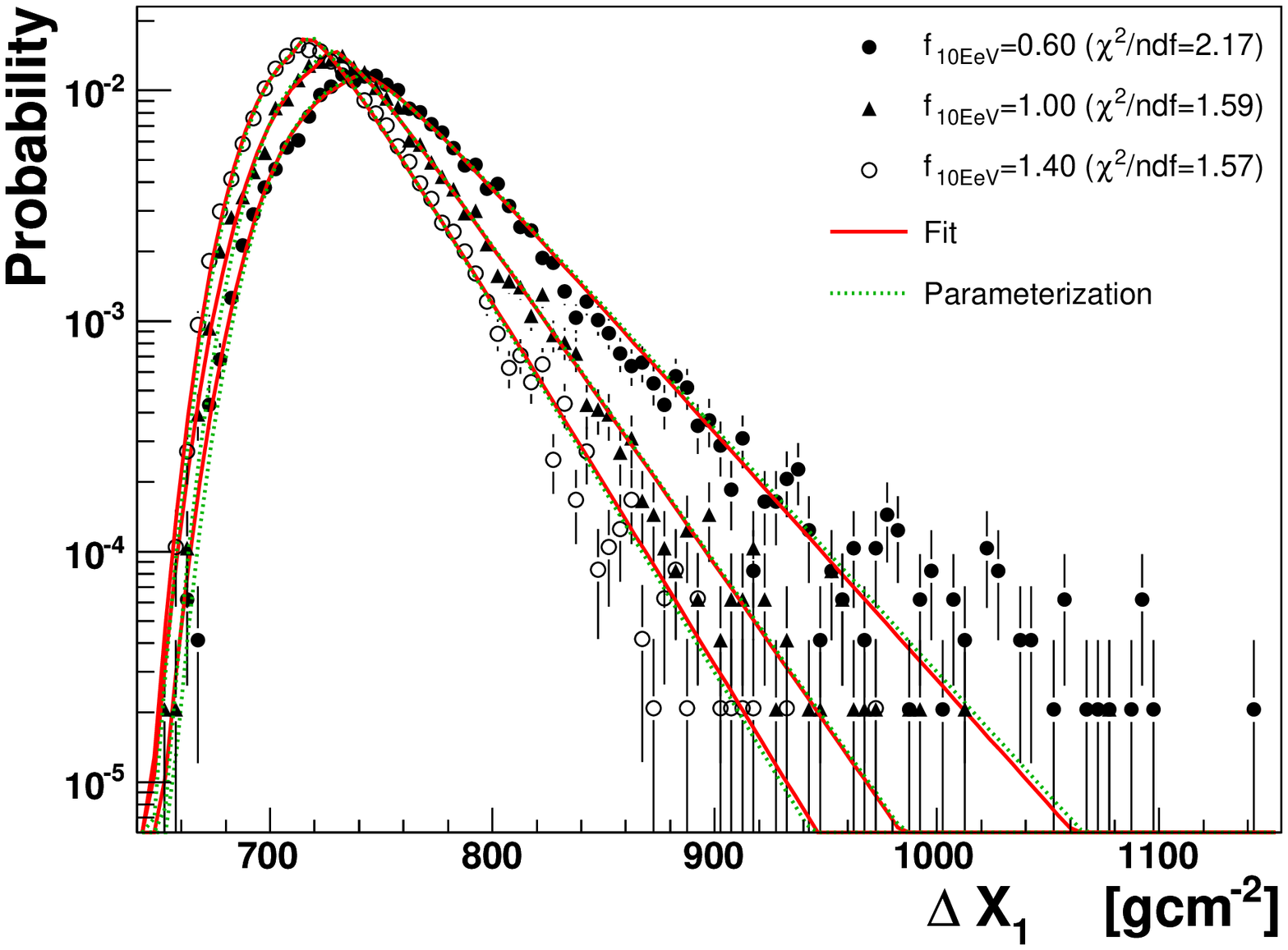}}~
  \hfill
  \subfigure[\textsc{SIBYLL 2.1}]{\includegraphics[width=.45\linewidth]{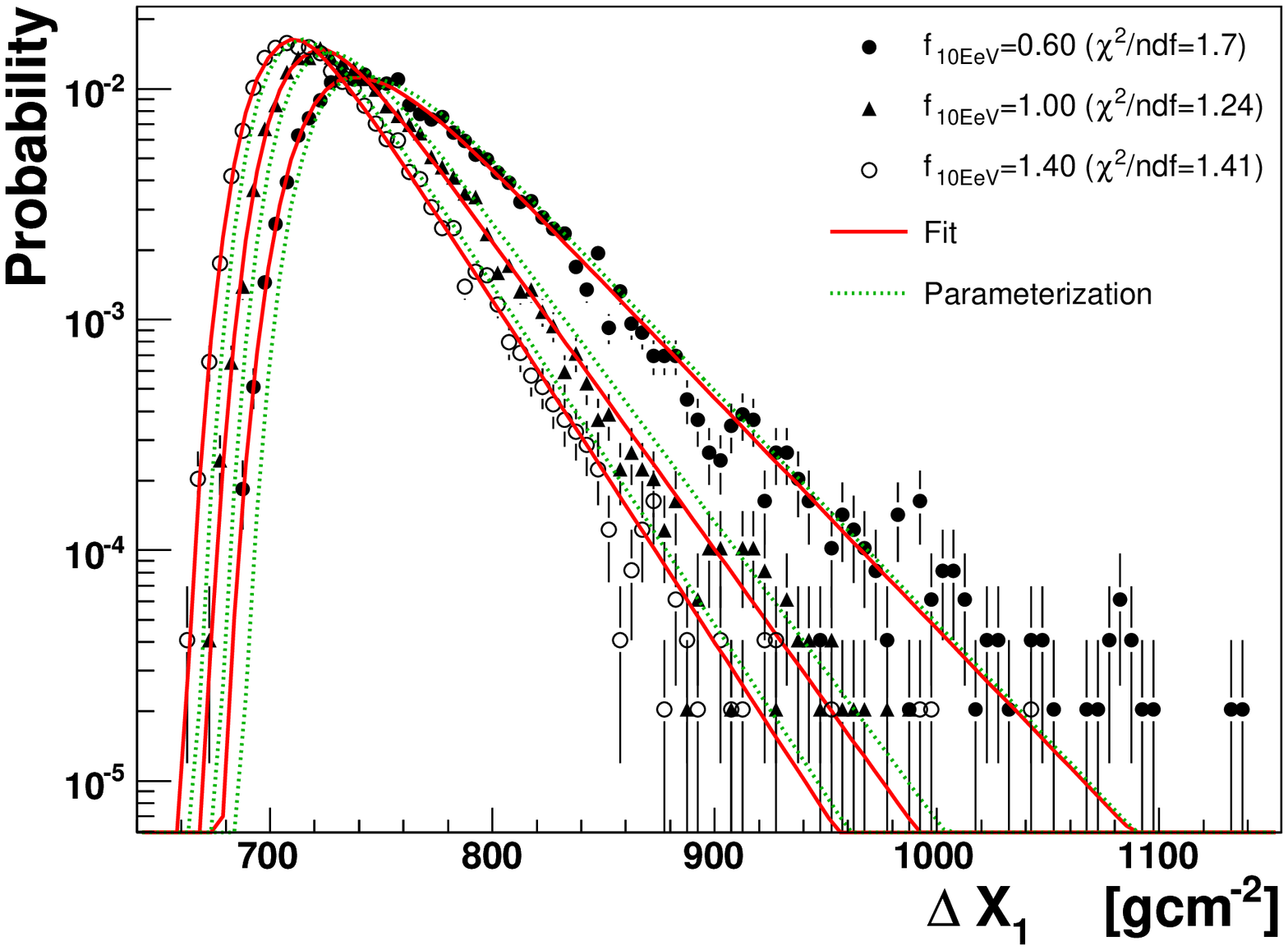}}~
  \hfill
  \caption{Simulated $\Delta X_{\rm 1}$-distributions at $10^{19}\,$eV
    for two hadronic interaction models. Included are the fits of the
    extended Moyal distribution Eq.~(\ref{eqn:moyalExt}) as well as
    the final parameterization.}
  \label{f:moyalDeltaX}
\end{figure*}
%%%%%%%%%%%%%%%%%%%%%%%%%%%%%%

\subsection{Fitting range and stability}
\label{sec:fittingrange}

To investigate the sensitivity of the method to the choice of the
$X_{\rm max}$-range used for the fit, we generate sets of 3000
simulated data showers and analyzes them with
Eq.~(\ref{eqn::improved}). The Gaussian detector resolution for
$X_{\rm max}$ is chosen to be \unit[20]{gcm$^{-2}$}, which is the
value reported by the Pierre Auger Collaboration~\cite{Dawson:2007di}.

%%%%%%%%%%%%%%%%%%%%%%%%%%%%%%%%
\begin{figure*}[bt!]
  \subfigure[Start of the fitting range, $X_{\rm max}^{\rm start}$ (relative to the peak of the $X_{\rm max}$-distribution)]{
    \includegraphics[width=.5\linewidth]{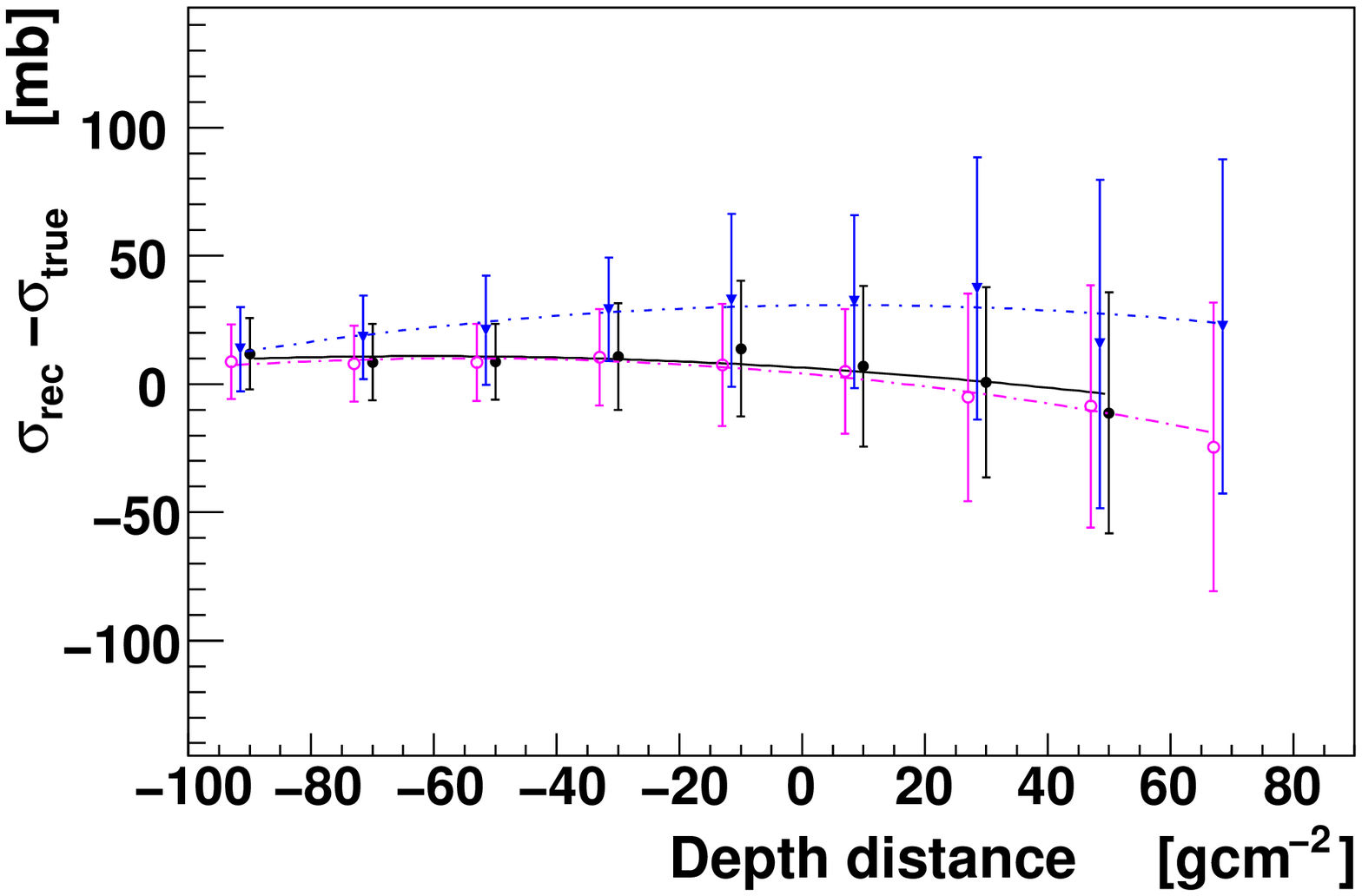}%
    \includegraphics[width=.5\linewidth]{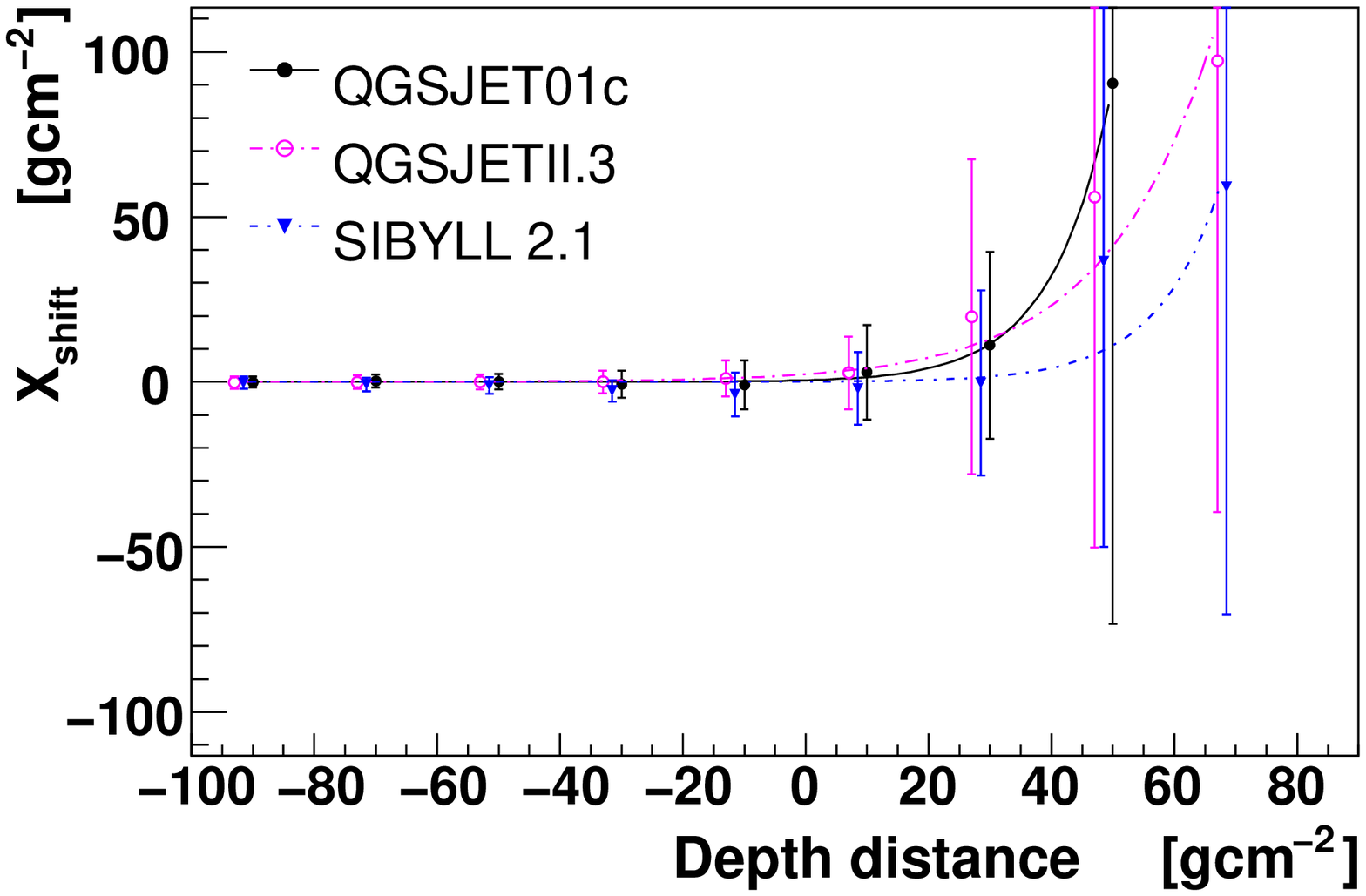}}\\
  \subfigure[End of the fitting range, $X_{\rm max}^{\rm stop}$  (relative to the peak of the $X_{\rm max}$-distribution)]{
    \includegraphics[width=.5\linewidth]{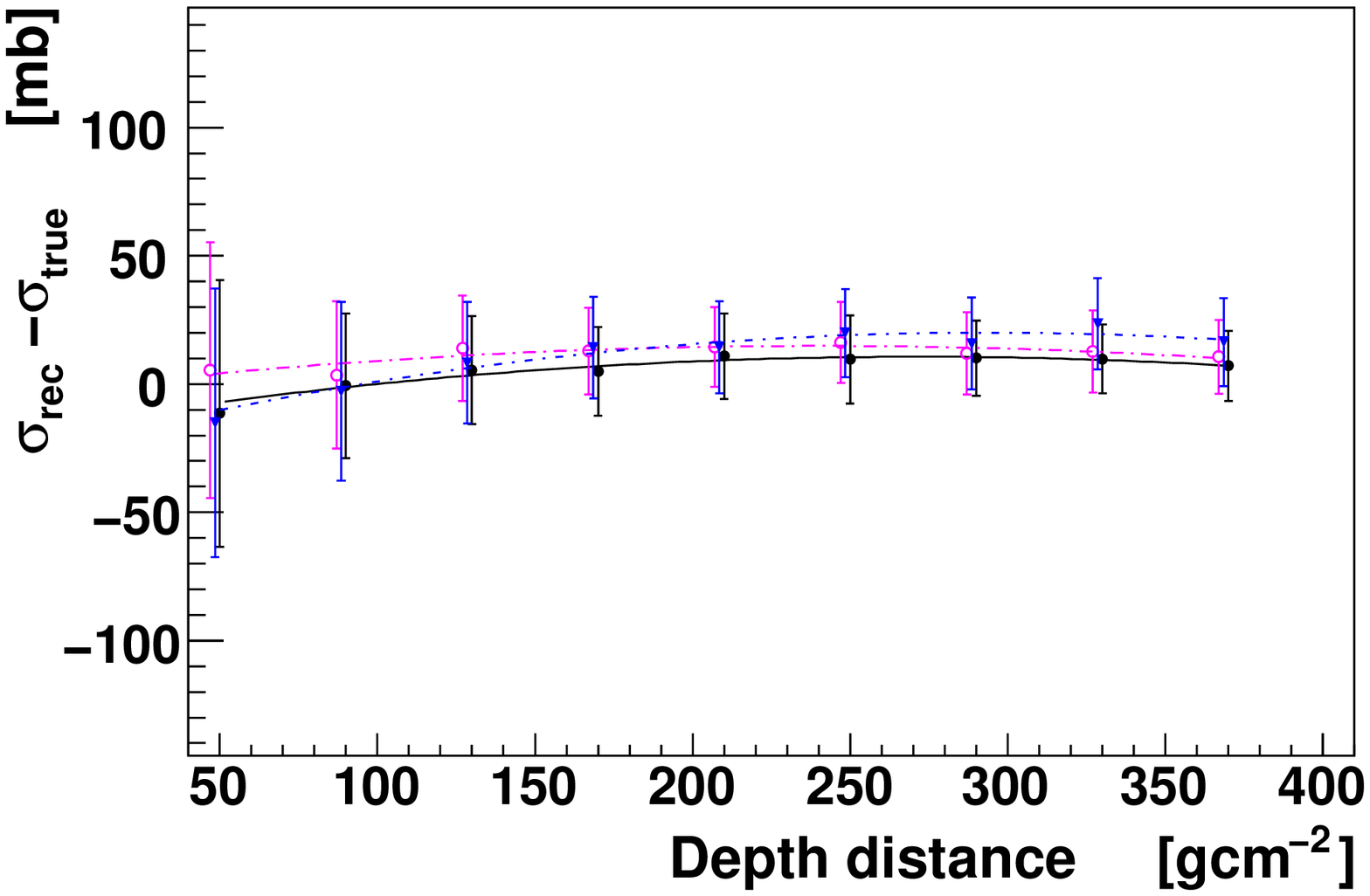}%
    \includegraphics[width=.5\linewidth]{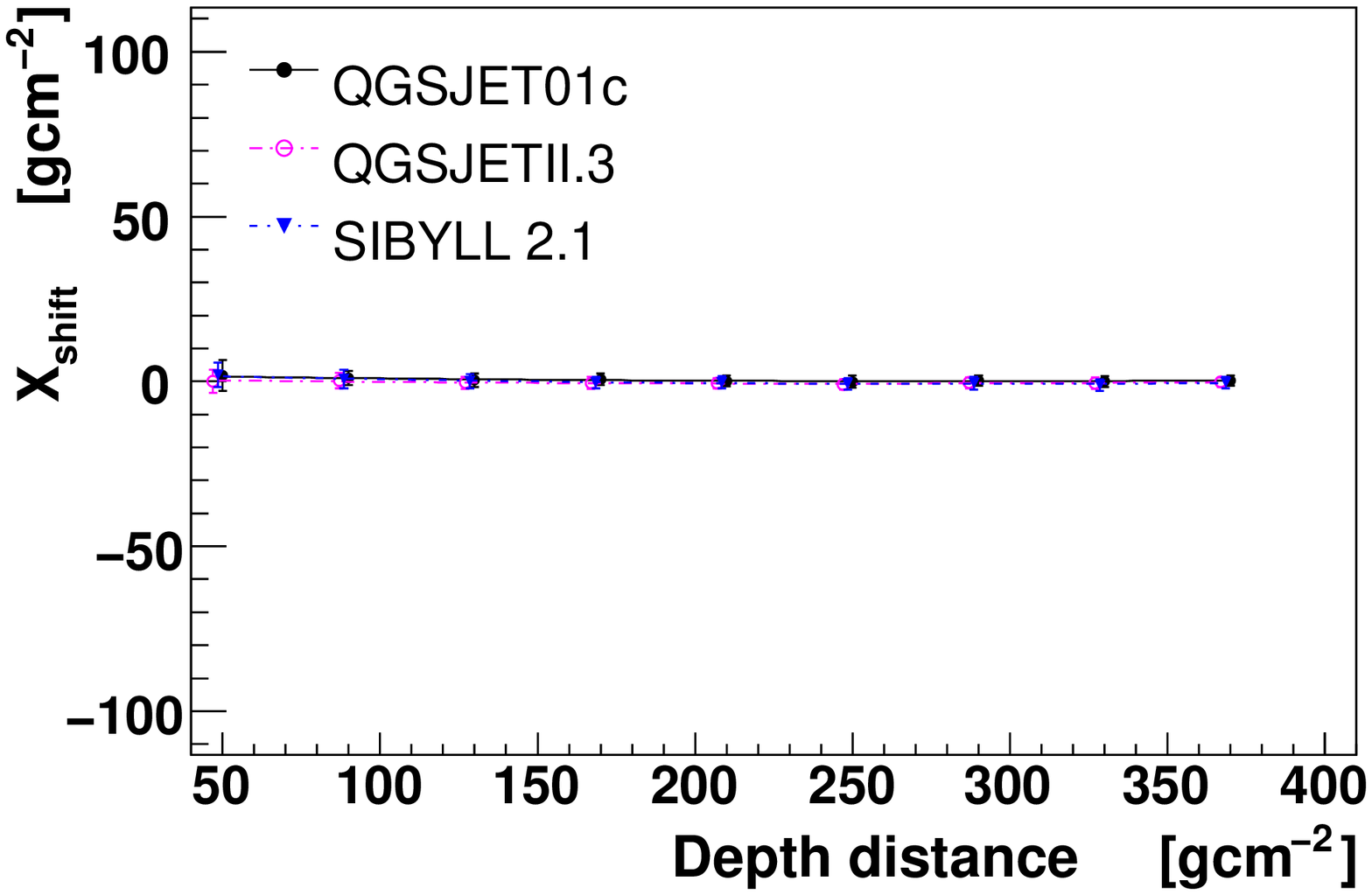}}
  \caption{Impact of the chosen fitting range in $X_{\rm max}^{\rm
      rec}$ on the resulting cross section (left) and $X_{\rm shift}$
    (right) at $\unit[10^{19}]{eV}$.  Each point corresponds to the mean of
    100 reconstructions and the error bars denote the resulting
    RMS. Both, the reconstructed $X_{\rm max}$ as well as the $\Delta
    X_1$-distribution are produced independently with the specified interaction
    model. The lines are just to guide the eye.}
  \label{f:FittingRange}
\end{figure*}
%%%%%%%%%%%%%%%%%%%%%%%%%%%%%%%%

The interval of the $X_{\rm max}$-distribution used to fit the model
is important for the resulting statistical as well as systematical
uncertainties. The first aspect is obvious since a reduction of the
dataset is clearly leading to a reduced statistical power of the
reconstruction.  The latter one is mostly because of the possible
contamination of the $X_{\rm max}$-distribution by cosmic ray
primaries other than protons.  All primary nuclei heavier than protons
are producing shallower $X_{\rm max}$ compared to protons. Primary
photons, on the other hand, are deeply penetrating and have a larger
$X_{\rm max}$ than protons.  A restricted fitting range in $X_{\rm
  max}$ can thus be used to enrich the considered fraction of protons
and reduce a possible contamination by other cosmic ray primaries.

The resulting impact of the chosen fitting range on the performance of
the reconstruction is shown in Figure~\ref{f:FittingRange}.  The
position of the peak of the $X_{\rm max}$-distribution, $X_{\rm
  peak}$, is used as a reference to define the fitting range. The
starting point as well as the ending point of the fit are expressed
only relative to $X_{\rm peak}$. 

Evidently, it is beneficial to chose the fitting range as large as
possible to get the smallest resulting statistical
uncertainties.  It is possible to shift the beginning of the
fitting range relatively close to $X_{\rm peak}$ without loosing too
much statistical power (cf. Figure~\ref{f:FittingRange}~(a)).  It is
found that, with the used parameterizations, the beginning of the
fitting range can be set to \unit[50]{gcm$^{-2}$} in front of $X_{\rm
  peak}$. In the following this is the adopted default choice.

The choice of the end of the fitting range has a
similar impact on the reconstruction. In
Figure~\ref{f:FittingRange}~(b) it is shown how the reconstruction is
degrading, while choosing a shorter fitting range.  Since the photon
fraction of ultra-high energy cosmic ray primaries is already strongly
constrained~\cite{Aglietta:2007yx} there is no special need to restrict
the upper end of the fitting range. In the following the upper end of the
fitting range for the log-likelihood fit of the $X_{\rm
  max}$-distribution is set to the value of the maximum
$X_{\rm max}$ plus \unit[40]{gcm$^{-2}$}.

The cross section dependent parameterizations of the $\Delta
X_1$-distributions are introducing a slight systematic overestimation
of the reconstructed cross sections of $\unit[10-20]{mb}$,
corresponding to $<5\%$. This can be taken into account within the
determination of the total systematic uncertainties of the
measurement. The statistical resolution of the reconstruction is
around $\unit[20]{mb}$ for 3000 events.

%%%%%%%%%%%%%%%%%%%%%%%%%%%%%%%%%%%%%%%%%%%%%%%%%%%%%%%%%%%%%%%%%

\section{Comparison of the performance of different analysis methods}

%%%%%%%%%%%%%%%%%%%%%%%%%%%%%%%%%%%%
\begin{figure}[p!]
  \centering
  \includegraphics[width=.75\textwidth]{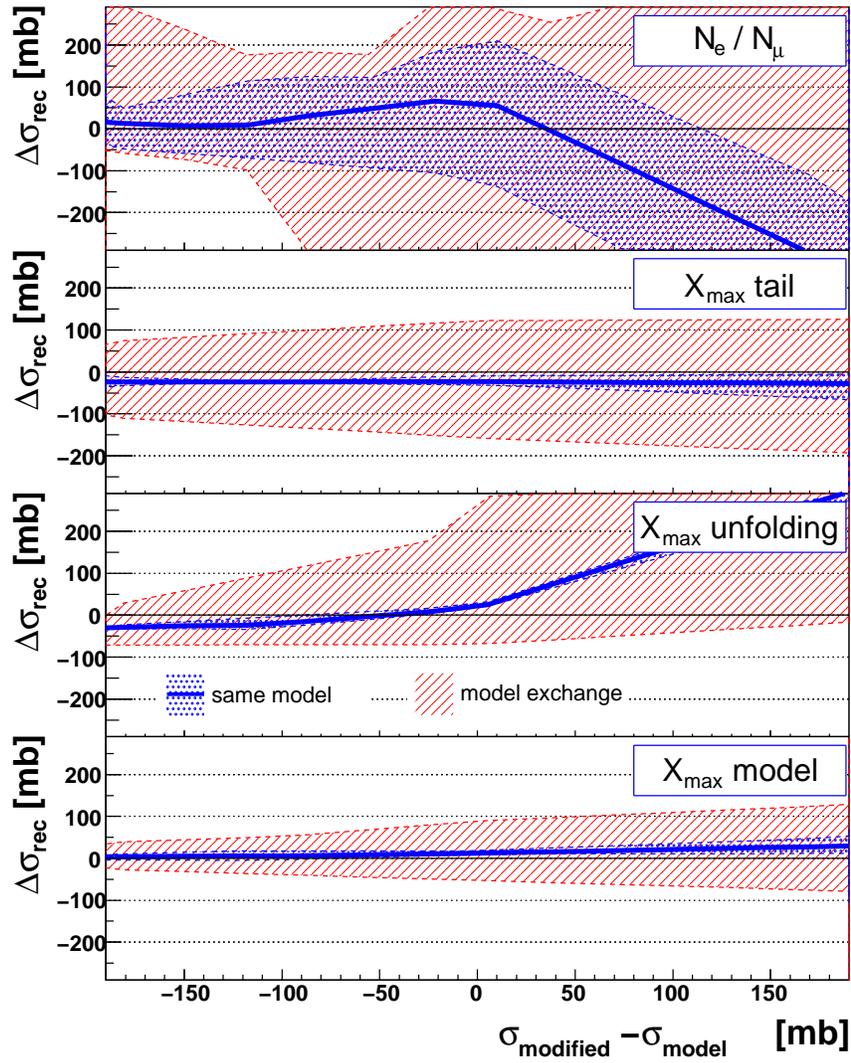}
  \vspace*{-0.8cm}
  \caption{Resulting systematics and dependence on models of cross
    section reconstruction techniques. Every analysis was performed
    100 times on 3000 simulated EAS; Shown are the mean results.  The
    central shaded area shows the ability of a method to reconstruct
    the input cross section of simulated air shower data with the same
    model that was used for the simulations. The thick line marks the
    mean result for all used models, which are \textsc{QGSJet01c},
    \textsc{QGSJetII.3} and \textsc{SIBYLL2.1}. The width of the
    central band is caused by statistical fluctuations. The outer
    shaded area denotes the maximum deviation of the cross section
    reconstructed with a model that differs from the one used to
    generated the simulated dataset. }
  \label{fig:Sensitivity}
\end{figure}
%%%%%%%%%%%%%%%%%%%%%%%%%%%%%%%%%%%%

In order to test the ability of a reconstruction method to recover a
changing input cross section, air shower simulations with a modified
cross section (cf.~\ref{sec:cxDep}) are performed.  The
simulated air shower data are then reconstructed and the found cross
sections $\sigma_{\rm rec}$ are compared to the modified input cross
sections $\sigma_{\rm modified}=f_{19}\,\sigma_{\rm model}$.
The result of this analysis for all cross section reconstruction
methods is summarized in Fig.~\ref{fig:Sensitivity}. The central shaded
area together with the solid line demonstrates the principal
self-consistency of the analysis technique. The outer shaded area can
be interpreted as the maximum possible model-dependence of the
reconstruction method. 

With the \NeNmu-method it is very difficult to reconstruct a modified cross
section. Additionally the model dependence of the result is generally
large. It is difficult to quantify, but can easily be of the order of
hundreds of mb.

A very much better performance is achieved by the $X_{\rm max}$-tail
analysis. The deviation of the reconstruction cross section from the
input value is not getting larger than $\unit[50]{mb}$ and the
model-dependence is between $\unit[100-200]{mb}$.

The $X_{\rm max}$-unfolding technique can find cross sections that are
not deviating much from the original model value. However, it shows a
clear systematic trend of underestimating small cross sections and
strongly overestimating large cross sections. Also the model
dependence is not negligible, ranging from $\unit[\sim75]{mb}$ at very
low cross sections, $\unit[\sim200]{mb}$ at intermediate cross sections
to many hundreds of mb at large cross sections.

Finally, the $X_{\rm max}$-model demonstrates its unique ability to retrieve
very consistently modified cross sections over a wide range. The model
dependence is smaller than for the $X_{\rm max}$-tail method and is
$\unit[<50]{mb}$ at low cross section while it grows to $\unit[\sim100]{mb}$ at
large cross sections.

The generally better sensitivity to reconstruct smaller compared to larger
cross section is resulting from the increasing importance of the fluctuations
of $X_1$ for the final $X_{\rm max}$-distribution. This facilitates the
measurement of a small cross section, while for large cross sections $X_{\rm
  max}$-distributions are mostly shaped by fluctuations of the shower
development $P_{\Delta X_1}$ and the detector resolution, making a measurement of the
cross section generally more difficult. 

%%%%%%%%%%%%%%%%%%%%%%%%%%%%%%%%%%%%%%%%%%%%%%
\begin{figure*}[bt!]
  \includegraphics[width=.499\textwidth]{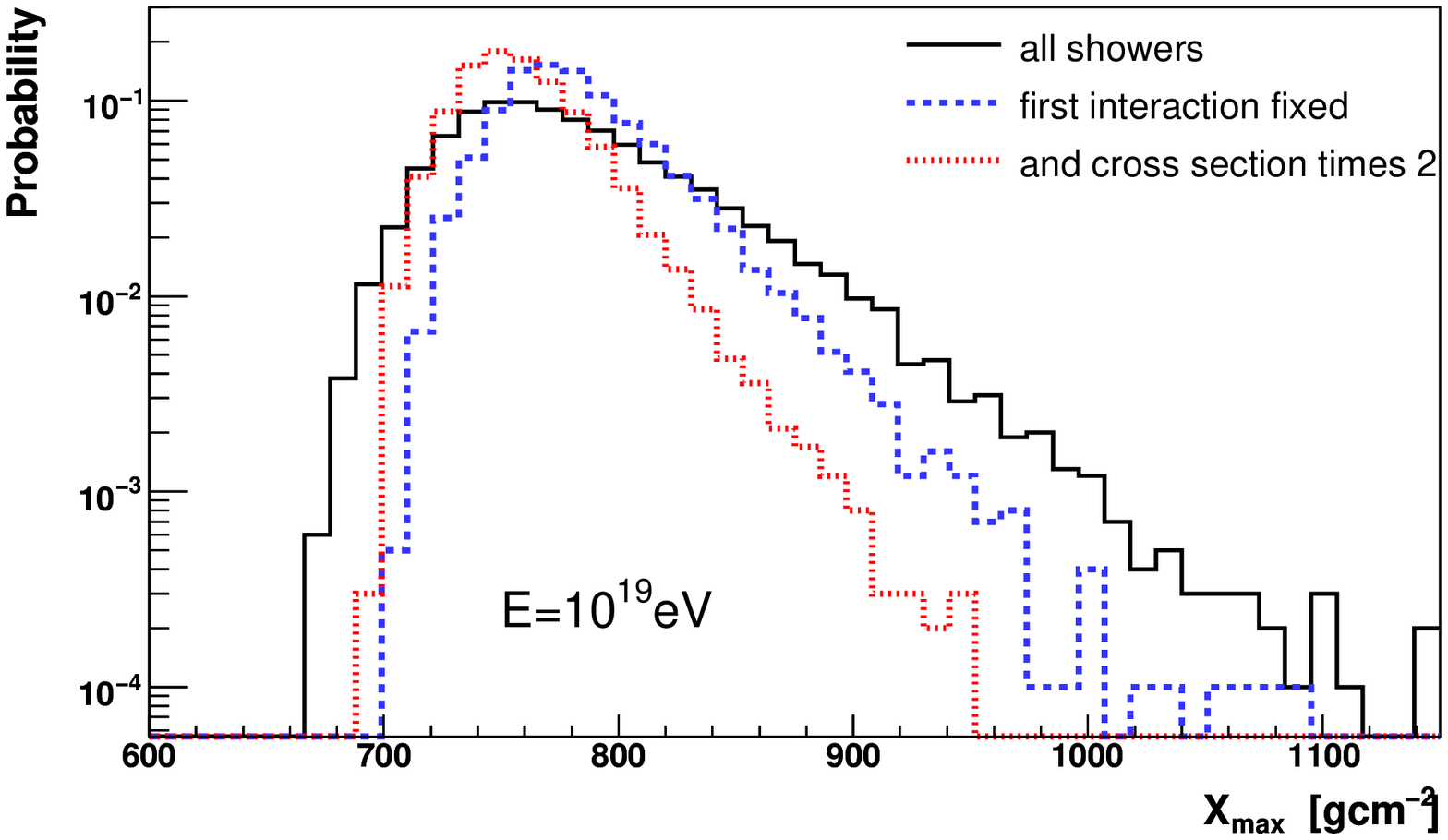}~
  \includegraphics[width=.499\textwidth]{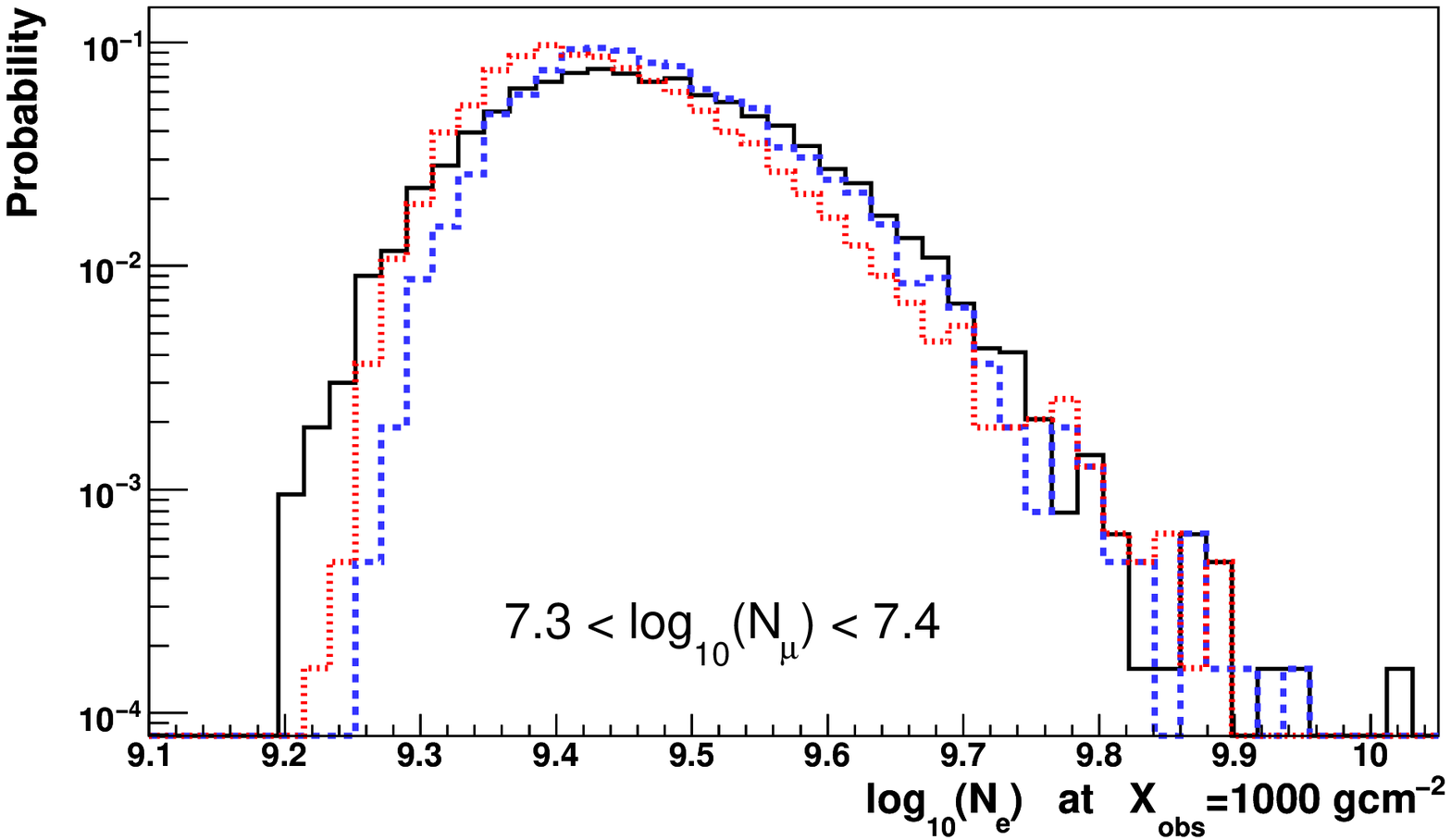}
  \caption{Shower-to-shower fluctuations of $X_{\rm max}$~(left) and
    $\log_{10}N_{\rm e}$~(right) for primary protons. The solid line
    is the original model prediction of \textsc{QGSJetII}, while the
    dashed line is for showers having their first interaction at the
    fixed slant depth corresponding to the predicted proton-air
    interaction length of \textsc{QGSJetII}. The dotted line is for
    simulated air showers with increased cross section by a factor of
    two, $f_{19}=2$, in addition to the fixed first
    interaction point.}
  \label{fig:ShowerFluctuations}
\end{figure*}
%%%%%%%%%%%%%%%%%%%%%%%%%%%%%%%%%%%%%%%%%%%%%%
Due to the combined impact of the fluctuation of both electron as well
as muon numbers, the remaining sensitivity of the \NeNmu-method to the
proton-air cross section is very much reduced compared to methods
using the observable $X_{\rm max}$, see
Fig.~\ref{fig:ShowerFluctuations}.

It is not straightforward to quantify the systematic effect caused by
interaction models on the results of the proton-air cross section
analyses. While the magnitude of the outer uncertainty band in
Fig.~\ref{fig:Sensitivity} suggests relatively large model dependence
of the results, this just reflects the existing incompatibility of the
underlying models. Thus, it cannot be used to make any conclusion on
the comparison of real data to the different models.  For the analysis
of real data, model dependence should be evaluated by the independent
reconstruction of the data based on all available interaction models,
and subsequent quoting of the mean result. The interval spanned by the
reconstructions can then be used as an estimation of the systematic
uncertainty induced by the models. This results in a smaller
dependence on interaction models compared to
Fig.~\ref{fig:Sensitivity}.

\section{Summary}

It is demonstrated in methodical studies that all techniques to
determine the proton-air cross section from air shower data are in
fact based on the same fundamental formulation that describes the
longitudinal air shower development. The requirement of using air
shower Monte Carlo simulations to interpret EAS observables inevitably
introduces a dependence on the hadronic interaction models used for
the simulations.

It is found that the magnitude of the model dependence is similar for
all cross section reconstruction methods; Only the \NeNmu-technique
exhibits a significantly larger model dependence, which is due to the
additional strong model dependence on the prediction of muon numbers.
In addition, since the prediction of electron and muon numbers are
both depending on interaction models, it is not possible to define
model independent cuts for the selection of air shower events.  Thus,
already on the level of event selection, a model dependence is
introduced in \NeNmu-analyses.

All $k$-factor based techniques are suffering from non-exponential
contributions to the slope of the $X_{\rm max}$- respectively
\NeNmu\ frequency-distribution at large depths. Thus the exponential
approximation used to define $k$-factors, is only of limited
accuracy. Any non-exponential contribution creates a strong dependence
of the exponential slope on the chosen fitting range~(see also
Ref.~\cite{AlvarezMuniz:2004bx}).  

Generally, $k$-factors are depending on the resolution of the
experiment and can therefore not be simply transferred from one
experiment to another, as done in some recent
analysis~\cite{Block:2007rq}.  In particular, $k_{\rm X}$-factors are
inherently different from $k_{\rm S}$-factors and can therefore not be
transferred from an $X_{\rm max}$-tail analysis to that of ground
based frequency attenuation or vice versa.  

%  We quantify this
%  effect to about $\unit[<10]{gcm^{-2}}$ for the $X_{\rm max}$-tail
%  but up to several tens of gcm$^{-2}$ for \NeNmu-method.  
% \item The values of all $k$-factors must be retrieved from Monte
%   Carlo simulations, which introduces a dependence on the interaction
%   models. This amounts to $\unit[\sim7]{\%}$ for $k_{\rm X}$, and to
%   $\unit[\sim28]{\%}$ for $k_{\rm S}$.
% \item It can not be disentangled whether a measurement of
%   $\Lambda_{\rm obs}$ can really be attributed to $\lambda_{\rm int}$
%   entirely or at least partly to changed fluctuations in $\Delta
%   X_{\rm 1}$ or $\Delta X_{\rm 2}$.

An $X_{\rm max}$-based analysis technique, as proposed in
Ref.~\cite{Belov:2006mb}, was further improved to reduce methodical
and model induced systematic uncertainties on the cross section result
as much as possible. The presented improved method is a model of the
$X_{\rm max}$-distribution based on simple considerations on
longitudinal air shower development combined with the parameterized
impact of a changing cross section on the resulting air shower
development.
Previous reconstruction techniques assume a static $\Delta
X_1$-distribution or $k$-factor, independent of the modified cross
section.  As, in general, the reconstructed cross section deviates
from the predicted value of the hadronic interaction model used to
generate the $\Delta X_1$-distribution or $k$-factor, a discrepancy of
the data with the air shower simulations is inevitable. The data can
not be described by the models used for the air shower simulations and
it is not sufficient to adopt the measured cross section just for the
first interaction point, while keeping the rest of the shower
simulations unchanged.

For the case of a proton-dominated cosmic ray data sample around
$\unit[10^{19}]{eV}$, which was studied in this article, it is
possible to determine the proton-air cross section with good
precision. The statistical uncertainties of the analysis of samples
with 3000 events are about \unit[20]{mb} (\unit[5]{\%}) and the
systematic uncertainties introduced by the parameterizations of the
$X_{\rm max}$-model are of the same order of magnitude.  The
model-induced systematic uncertainty is between 50 and \unit[100]{mb},
corresponding to relative uncertainties of $\unit[<10]{\%}$ to
$\unit[\sim20]{\%}$. A further reduction of this model-related
uncertainty is possible, but needs improvements in the understanding
of the underlying hadronic interaction physics.

Furthermore, the second free parameter of the new analysis technique,
$X_{\rm shift}$, is a further handle that is sensitive to the physics
of hadronic interactions at ultra-high energies.  A non-zero result of
$X_{\rm shift}$ indicates the deficiency of the underlying hadronic
interaction model to describe the measured data distribution. It can
be interpreted in terms of a modified characteristics of secondary
particle production in hadronic interactions~\cite{Ulrich:2009xxx}.

\ack The authors thank their colleagues from the Pierre Auger
Collaboration for helpful discussions on methods for measuring the
proton-air cross section. This work has been supported in part by
Bundesministerium f{\"u}r Bildung und Forschung (BMBF) grant No.\
05A08VK1.

\clearpage
\begin{appendix}
\section{Parameterization of $\Delta X_{\rm 1}$-distribution}
\label{appendix}

A modified version of the Moyal distribution is introduced to
parameterize the $\Delta X_{\rm 1}$-distribution
\begin{equation}
  \label{eqn:moyalExt}
  p_{\rm 1}(\Delta X_{\rm 1})=\frac{N}{\beta\,\sqrt{2\pi}} \;
  e^{-\frac{1}{2}\left(z+e^{-z\;\left(z^2\right)^\gamma}\right)}
  \qquad \mbox{with}\qquad z=\frac{\Delta X_{\rm 1}-\alpha}{\beta}\,.
\end{equation}
The additional third degree of freedom $\gamma$ makes the
parameterization more flexible, allowing us to improve the description
around the peak of the $\Delta X_{\rm 1}$-distributions.  The
normalization value $N$ is not a free parameter, but is numerically computed to correctly normalize the
distribution. Since the Moyal function with the two parameters
$\alpha$ and $\beta$ itself is normalized, we only have to correct for
the $\gamma$ parameter
\begin{equation}
  N(\gamma)=\frac{1}{0.965 + 3.685\times10^{-3}\; \bigl(\gamma +
    0.366\bigr)^{-2.171}} \,.
\end{equation}

\subsection{Energy dependence}
\label{sec:DeltaX}

To determine the energy dependence of the parameters $\alpha$, $\beta$
and $\gamma$, the parameterization Eq.~(\ref{eqn:moyalExt}) is fitted
to $\Delta X_{\rm 1}$-distributions generated by CONEX at several
primary energies. The energy dependence can then be interpolated by a
polynomial of 2$^{\rm nd}$ degree
\begin{eqnarray}
  \label{eqn:pol2edep}
  \alpha^{\rm e}(E) &= \alpha_0^{\rm e} + \alpha_1^{\rm e} \;
  \log_{\rm 10}(E/\mbox{eV}) + \alpha_2^{\rm e} \; \log_{\rm
    10}^2(E/\mbox{eV}) \nonumber\\ \beta^{\rm e}(E) &= \beta_0^{\rm e}
  + \beta_1^{\rm e} \; \log_{\rm 10}(E/\mbox{eV}) + \beta_2^{\rm e}
  \; \log_{\rm 10}^2(E/\mbox{eV}) \\ \gamma^{\rm e}(E) &=
  \gamma_0^{\rm e} + \gamma_1^{\rm e} \; \log_{\rm 10}(E/\mbox{eV})
  + \gamma_2^{\rm e} \; \log_{\rm 10}^2(E/\mbox{eV}) \nonumber\,.
\end{eqnarray}

\begin{table}[b!]
  \centering 
  \caption{Parameters for describing the energy dependence of $\Delta X_1$-distributions.} 
  \footnotesize
  \vspace*{.25cm}
  \begin{tabular}{l@{\hspace*{.2cm}\vrule width 1.1pt\hspace*{.2cm}}c|r@{$\pm$}lr@{$\pm$}lr@{${\pm}$}l} 
    \textbf{Model} & \textbf{\boldmath Index $i$} &
    \multicolumn{2}{c}{\boldmath\bf$\alpha^{\rm e}_i$ [g/cm]} &
    \multicolumn{2}{c}{\boldmath\bf$\beta^{\rm e}_i$ [g/cm]} &
    \multicolumn{2}{c}{\boldmath\bf$\gamma^{\rm e}_i$ [1]} \\ \hline
    \hline
     \multirow{3}{*}{\textsc{QGSJet01c}} & $0$ &-871.25 & 27.76 & 85.15 & 14.49 & 5.94 & 1.43 \\
  & $1$ &113.91 & 3.01 & -7.49 & 1.57 & -0.71 & 0.16 \\
  & $2$ &-1.64 & 0.08 & 0.22 & 0.04 & 0.02 & 0.00 \\
      \hline
     \multirow{3}{*}{\textsc{QGSJetII.3}} & $0$ &-1320.67 & 24.86 & 226.24 & 11.68 & 6.04 & 0.39 \\
  & $1$ &165.29 & 2.70 & -22.31 & 1.27 & -0.63 & 0.04 \\
  & $2$ &-3.04 & 0.07 & 0.59 & 0.03 & 0.02 & 0.00 \\
      \hline
     \multirow{3}{*}\textsc{{SIBYLL 2.1}} & $0$ &-606.32 & 20.37 & 131.27 & 9.56 & 0.33 & 0.43 \\
  & $1$ &79.17 & 2.20 & -11.78 & 1.03 & -0.05 & 0.05 \\
  & $2$ &-0.45 & 0.06 & 0.30 & 0.03 & 0.00 & 0.00 \\
  \end{tabular}
 \label{t:edepX1}
\end{table}

\begin{figure*}[t!]
  \includegraphics[width=\linewidth]{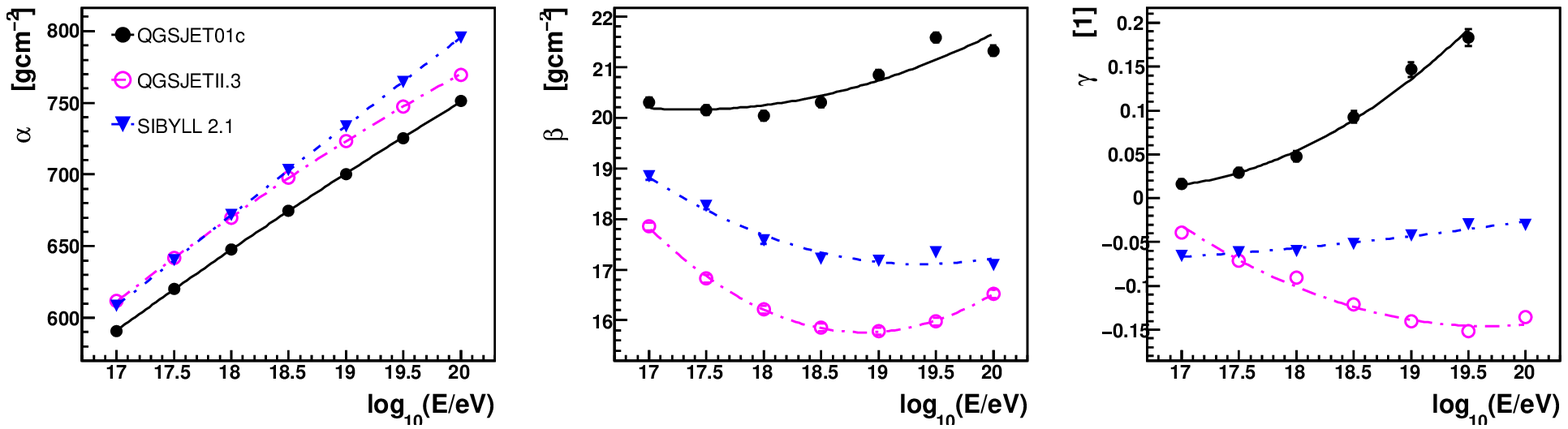}
  \caption{Parameterization of the energy dependence of the $\Delta
    X_{\rm 1}$-distributions.  The lines denote the polynomial
    interpolations given in the text.}
  \label{f:CXenergy}
\end{figure*}

The chosen polynomial interpolation is well suited to reproduce the
found dependences over the considered energy range.  The results are
illustrated in Fig.~\ref{f:CXenergy} and the parameters are listed in
Table~\ref{t:edepX1}. 
A first result is that even the energy dependence of $\Delta X_{\rm
  1}$ is depending on the high energy interaction model used during
air shower simulations.

\subsection{Cross section dependence}
\label{sec:cxDep}

To infer the dependence of $P_{\Delta X_1}$ on $\lambda_{\rm p-air}$,
a modified CONEX version~\cite{Ulrich:2009xxx} was used to generate
$\Delta X_1$-distributions for several values of $f_{19}$, see
Eq.~(\ref{eq:rescalingFactor}).  The energy dependence of all hadronic
cross sections, i.e.\ p-, $\pi$-, and $K$-air interactions are then
altered by the modification factor.  This implies that the model
uncertainties above $\unit[10^{15}]{eV}$ are rising proportional to
the logarithm of the energy.  This choice seems certainly reasonable,
however, in the future also other energy dependences of $f(E)$ may
turn out to be useful.

\begin{table}[b!]
  \centering
  \caption{Parameters for describing the cross section dependence of
    $\Delta X_1$-distributions.}
  \vspace*{.25cm}
  \footnotesize
  \begin{tabular}{l@{\hspace*{.2cm}\vrule width 1.1pt\hspace*{.2cm}}c|r@{$\pm$}lr@{$\pm$}lr@{${\pm}$}l} 
    \textbf{Model} & \textbf{\boldmath Index $i$} &
    \multicolumn{2}{c}{\boldmath\bf$\alpha^{\sigma}_i$ [g/cm]} &
    \multicolumn{2}{c}{\boldmath\bf$\beta^{\sigma}_i$ [g/cm]} &
    \multicolumn{2}{c}{\boldmath\bf$\gamma^{\sigma}_i$ [1]} \\ \hline
    \hline
     \multirow{3}{*}{\textsc{QGSJet01c}} & $0$ &730.09 & 0.25 & 10.33 & 0.00 & 0.03 & 0.03 \\
  & $1$ &-33.61 & 0.34 & 11.83 & 0.00 & 0.10 & 0.03 \\
  & $2$ &5.07 & 0.09 & -0.13 & 0.00 & 0.20 & 0.07 \\
  & $3$ &  \multicolumn{4}{c}{~}  & -0.01 & 0.01 \\
      \hline
     \multirow{3}{*}\textsc{{QGSJetII.3}} & $0$ &754.10 & 0.27 & 8.46 & 0.08 & -0.25 & 0.02 \\
  & $1$ &-35.86 & 0.35 & 8.28 & 0.19 & 0.10 & 0.02 \\
  & $2$ &5.14 & 0.09 & -0.08 & 0.01 & -0.20 & 0.09 \\
  & $3$ &  \multicolumn{4}{c}{~} & 0.02 & 0.00 \\
      \hline
     \multirow{3}{*}\textsc{{SIBYLL 2.1}} & $0$ &774.36 & 0.27 & 11.67 & 0.09 & -0.18 & 0.04 \\
  & $1$ &-46.36 & 0.36 & 4.76 & 0.16 & 0.10 & 0.04 \\
  & $2$ &7.12 & 0.10 & 0.12 & 0.01 & -0.00 & 0.15 \\
  & $3$ & \multicolumn{4}{c}{~} & 0.04 & 0.01 \\
  \end{tabular}
 \label{t:cxdepX1}
\end{table}
\begin{figure*}[t!]
  \includegraphics[width=\linewidth]{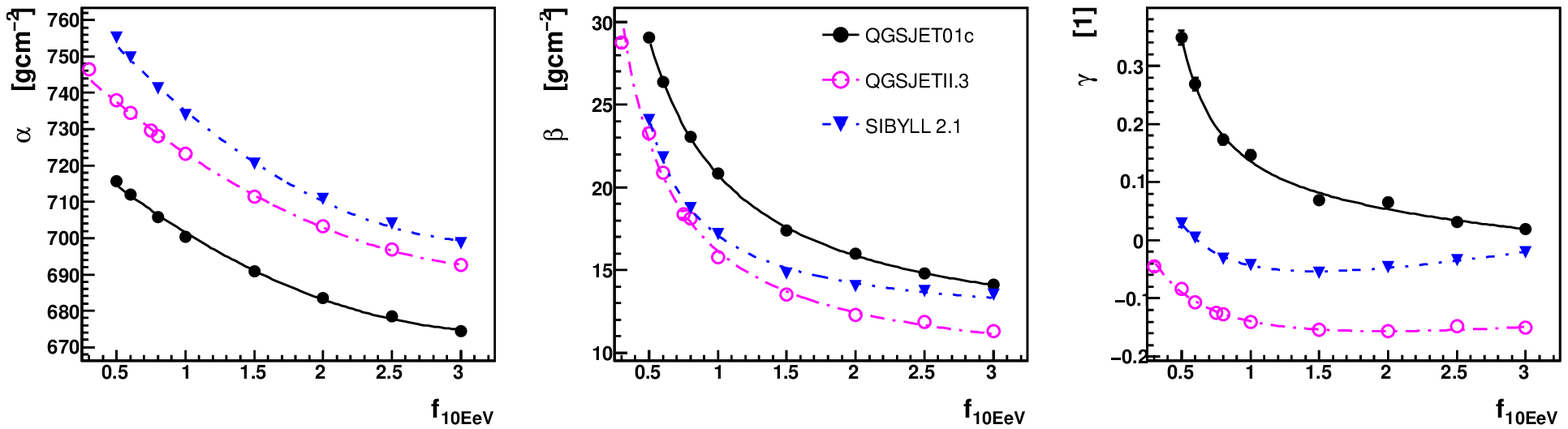}
  \caption{Parameterization of the $f_{19}$-dependence of the
    $\Delta X_{\rm 1}$-parameterizations with respect to a changing
    cross section.  The lines denote the interpolations as
    discussed in the text.}
  \label{f:CXdeltaX}
\end{figure*} 

The resulting dependence of $\Delta X_{\rm 1}$ on a changing cross
section for air showers at $\unit[10^{19}]{eV}$ is shown in
Fig.~\ref{f:CXdeltaX}.

  The dependence of the $\alpha$, $\beta$ and
$\gamma$ parameters of the modified Moyal distribution on $f_{19}$ is parameterized as follows
\begin{eqnarray}
  \label{eqn:pol2cxdep}
  \alpha^{\sigma}(f_{19}) &= \alpha_0^{\sigma} + \alpha_1^{\sigma} \; f_{19}  + \alpha_2^{\sigma} \; f_{19}^2\nonumber\\
  \beta^{\sigma}(f_{19})  &= \beta_0^{\sigma} + \beta_1^{\sigma} / (f_{19} - \beta_2^{\sigma})\\
  \gamma^{\sigma}(f_{19}) &= \gamma_0^{\sigma} + \gamma_1^{\sigma} / (f_{19} - \gamma_2^{\sigma}) + \gamma_3^{\sigma} \; f_{19}\nonumber\;\;.
\end{eqnarray}
The resulting parameters of the interpolations are listed in Table~\ref{t:cxdepX1}. 
Furthermore, it is convenient to introduce the relative change as 
\begin{eqnarray}
  \label{eqn:pol2cxdepRel}
  \Delta\alpha^{\sigma}(f_{19}) &= \alpha^{\sigma}(f_{19})-\alpha^{\sigma}(1) \nonumber\\
  \Delta\beta^{\sigma}(f_{19})  &= \beta^{\sigma}(f_{19})-\beta^{\sigma}(1)\\
  \Delta\gamma^{\sigma}(f_{19}) &= \gamma^{\sigma}(f_{19})-\gamma^{\sigma}(1)\nonumber \;\;.
\end{eqnarray}

\subsection{Parameterization of the $\Delta X_{\rm 1}$-distribution}

Combining the found dependence of $\Delta X_{\rm 1}$-distributions on
the energy $E$ and the cross section modifier $f_{19}$, it is
now possible to construct a parameterization of $P_{\Delta X_1}$ in terms of $E$
and $\sigma(E)$.  Firstly, the principle energy dependence of $P_{\Delta X_1}$ is
evaluated by calculating $\alpha^{\rm e}(E)$, $\beta^{\rm e}(E)$ and
$\gamma^{\rm e}(E)$ using Eq.~(\ref{eqn:pol2edep}) and the results
from Tab.~\ref{t:edepX1}.  Secondly, the effect of the changed cross
section is added. To achieve this, the first step is to calculate the
corresponding deviation of the cross section $\sigma(E)$ extrapolated
to $\unit[10^{19}]{eV}$ assuming Eq.~(\ref{eq:rescalingFactor}).
The resulting factor
$f_{19}$ is then used to evaluate $\Delta\alpha^{\sigma}(f_{19})$, $\Delta\beta^{\sigma}(f_{19})$ and
$\Delta\gamma^{\sigma}(f_{19})$ using
Eq.~(\ref{eqn:pol2cxdepRel}) and the results listed in
Tab.~\ref{t:cxdepX1}.  In accordance to Eq.~(\ref{eq:rescalingFactor}) the
energy dependence is assumed to be logarithmic and vanishing
below $\unit[10^{15}]{eV}$.  This yields the final set of parameters of $P_{\Delta X_1}$
\begin{eqnarray}
  \label{eqn:deltaXparam}
  \alpha\bigl(E,\,\sigma(E)\bigr) &= \alpha^{\rm e}\bigl(E\bigr)\;+\; F\bigl(E\bigr) \;
  \Delta\alpha^{\sigma}\bigl(f_{19}\bigr)\nonumber\\ \beta\bigl(E,\,\sigma(E)\bigr) &=
  \beta^{\rm e}\bigl(E\bigr)\;+\;F\bigl(E\bigr) \; \Delta\beta^{\sigma}\bigl(f_{19}\bigr)\\ \gamma\bigl(E,\,\sigma(E)\bigr) &= \gamma^{\rm
    e}\bigl(E\bigr)\;+\;F\bigl(E\bigr) \; \Delta\gamma^{\sigma}\bigl(f_{19}\bigr)\nonumber\,,
\end{eqnarray}
with
\begin{equation}
  F(E) = \left\{
  \begin{array}{l l}
    0 & \quad E\le\unit[10^{15}]{eV}\\
    \ln(E/1\,\mbox{PeV})/\ln(10\,\mbox{EeV}/1\,\mbox{PeV}) & \quad 
    E>\unit[10^{15}]{eV}
  \end{array}\right. \,.
\end{equation}

\end{appendix}
 
% --------------------------------------------------------------------------
% Bibliography
% --------------------------------------------------------------------------
\clearpage
\section*{Bibliography}
\bibliographystyle{unsrt-mod-notitle}
\bibliography{article}

\begin{thebibliography}{10}

\bibitem{Bodhaine:1999}
B.~Bodhaine {\it et~al.},
J. Atmos. Ocean. Tech. 16 (1999) 1854.

\bibitem{Grigorov:1965aa}
N.~Grigorov {\it et~al.},
Proc. of 9$^{\rm th}$ Int. Cosmic Ray Conf, London 1 (1965) 860.

\bibitem{Yodh:1972fv}
G.~Yodh, Y.~Pal, and J.~Trefil,
Phys. Rev. Lett. 28 (1972) 1005--1008.

\bibitem{Nam:1975aa}
R.~Nam, S.~Nikolsky, V.~Pavluchenko, A.~Chubenko, and V.~Yakovlev,
{Proc. of 14$^{\rm th}$ Int. Cosmic Ray Conf}, Munich  (1975) 2258.

\bibitem{Siohan:1978zk}
F.~Siohan {\it et~al.},
J. Phys. G4 (1978) 1169--1186.

\bibitem{Mielke94}
H.~Mielke, M.~Foeller, J.~Engler, and J.~Knapp,
J. Phys. G20 (1994) 637--649.

\bibitem{Hara:1983pa}
T.~Hara {\it et~al.},
Phys. Rev. Lett. 50 (1983) 2058--2061.

\bibitem{Honda:1993kv}
M.~Honda {\it et~al.},
Phys. Rev. Lett. 70 (1993) 525--528.

\bibitem{Nikolaev:1993mc}
N.~Nikolaev,
Phys. Rev. D48 (1993) 1904--1906
 and hep-ph/9304283.

\bibitem{Aglietta:2009zz}
M.~Aglietta {\it et~al.},
Phys. Rev. D79 (2009) 032004.

\bibitem{Baltrusaitis:1984ka}
R.~Baltrusaitis {\it et~al.},
Phys. Rev. Lett. 52 (1984) 1380--1383.

\bibitem{Knurenko:1999cr}
S.~Knurenko, V.~Sleptsova, I.~Sleptsov, N.~Kalmykov, and S.~Ostapchenko,
{Proc. of 26$^{\rm th}$ Int. Cosmic Ray Conf.}, Salt Lake City, Utah 1 (1999)
  372.

\bibitem{Belov:2006mb}
K.~Belov,
Nucl. Phys. (Proc. Suppl.) 151 (2006) 197--204.

\bibitem{Collaboration:2009ca}
G.~Aielli {\it et~al.},
submitted to Phys. Rev. D  (2009)
 and arXiv:0904.4198 [hep-ex].

\bibitem{Gaisser:1993ix}
T.~K. Gaisser {\it et~al.}  (HIRES Collab.),
Phys. Rev. D47 (1993) 1919--1932.

\bibitem{Kalmykov:1997te}
N.~Kalmykov, S.~Ostapchenko, and A.~Pavlov,
Nucl. Phys. (Proc. Suppl.) B52 (1997) 17--28.

\bibitem{Ostapchenko:2005nj}
S.~Ostapchenko,
Phys. Rev. D74 (2006) 014026
 and hep-ph/0505259.

\bibitem{Engel:1999db}
R.~Engel, T.~Gaisser, T.~Stanev, and P.~Lipari,
{Proc. of 26$^{\rm th}$ Int. Cosmic Ray Conf.}, Salt Lake City, Utah 1 (1999)
  415.

\bibitem{Fletcher:1994bd}
R.~Fletcher, T.~Gaisser, P.~Lipari, and T.~Stanev,
Phys. Rev. D50 (1994) 5710--5731.

\bibitem{Werner:2007vd}
K.~Werner and T.~Pierog,
AIP Conf. Proc. 928 (2007) 111--117
 and arXiv:0707.3330 [astro-ph].

\bibitem{Ostapchenko:2004ss}
S.~Ostapchenko,
Nucl. Phys. Proc. Suppl. 151 (2006) 143--146
 and hep-ph/0412332.

\bibitem{Engel:1998pw}
R.~Engel, T.~Gaisser, P.~Lipari, and T.~Stanev,
Phys. Rev. D58 (1998) 014019
 and hep-ph/9802384.

\bibitem{Berezinsky:2005cq}
V.~Berezinsky, A.~Gazizov, and S.~Grigorieva,
Phys. Lett. B612 (2005) 147--153
 and astro-ph/0502550.

\bibitem{Berezinsky:2006nq}
V.~Berezinsky, S.~Grigoreva, and B.~Hnatyk,
Nucl. Phys. (Proc. Suppl.) 151 (2006) 497--500.

\bibitem{Berezinsky:2007wf}
V.~Berezinsky,
{Proc. of 30$^{\rm th}$ Int. Cosmic Ray Conf.}, Merida, Mexico  (2007)
 and arXiv:0710.2750 [astro-ph].

\bibitem{Zatsepin:1966jv}
G.~Zatsepin and V.~Kuzmin,
JETP Lett. 4 (1966) 78--80.

\bibitem{Greisen:1966jv}
K.~Greisen,
Phys. Rev. Lett. 16 (1966) 748--750.

\bibitem{Cronin:2007zz}
J.~Abraham {\it et~al.}  (Pierre Auger Collab.),
Science 318 (2007) 938--943
 and arXiv:0711.2256 [astro-ph].

\bibitem{Abbasi:2007sv}
R.~Abbasi {\it et~al.}  (HiRes Collab.),
Phys. Rev. Lett. 100 (2008) 101101
 and astro-ph/0703099.

\bibitem{Abraham:2008ru}
J.~Abraham {\it et~al.}  (Pierre Auger Collab.),
Phys. Rev. Lett. 101 (2008) 061101
 and arXiv:0806.4302 [astro-ph].

\bibitem{Ulrich:2009xxy}
R.~Ulrich {\it et~al.},
in preparation  (2009) .

\bibitem{Abbasi:2004nz}
R.~Abbasi {\it et~al.}  (HiRes Collab.),
Astrophys. J. 622 (2005) 910--926
 and astro-ph/0407622.

\bibitem{Abraham:2004dt}
J.~Abraham {\it et~al.}  (Pierre Auger Collab.),
Nucl. Instrum. Meth. A523 (2004) 50.

\bibitem{Kawai:2008zza}
H.~Kawai {\it et~al.}  (TA Collab.),
Nucl. Phys. Proc. Suppl. 175-176 (2008) 221--226.

\bibitem{Landau:1953um}
L.~Landau and I.~Pomeranchuk,
Dokl. Akad. Nauk Ser. Fiz. 92 (1953) 535--536.

\bibitem{Landau:1953gr}
L.~Landau and I.~Pomeranchuk,
Dokl. Akad. Nauk Ser. Fiz. 92 (1953) 735--738.

\bibitem{Migdal:1956tc}
A.~Migdal,
Phys. Rev. 103 (1956) 1811--1820.

\bibitem{Bergmann:2006yz}
T.~Bergmann {\it et~al.},
Astropart. Phys. 26 (2007) 420--432
 and astro-ph/0606564.

\bibitem{AlvarezMuniz:2002xs}
J.~Alvarez-Muniz, R.~Engel, T.~Gaisser, J.~Ortiz, and T.~Stanev,
Phys. Rev. D66 (2002) 123004
 and astro-ph/0209117.

\bibitem{Rossi:1941zz}
B.~Rossi and K.~Greisen,
Rev. Mod. Phys. 13 (1941) 240--309.

\bibitem{Billoir:2005}
{P. Billoir},
\url{http://www-ik.fzk.de/corsika/corsika-school2008/talks/}
CORSIKA School, 2008.

\bibitem{Nerling:2006yt}
F.~Nerling, J.~Bl{\"u}mer, R.~Engel, and M.~Risse,
Astropart. Phys. 24 (2006) 421--437.

\bibitem{Giller:2005qz}
M.~Giller, A.~Kacperczyk, J.~Malinowski, W.~Tkaczyk, and G.~Wieczorek,
J. Phys. G31 (2005) 947--958.

\bibitem{Lafebre:2009en}
S.~Lafebre {\it et~al.},
Astropart. Phys. 31 (2009) 243--254
 and arXiv:0902.0548 [astro-ph.HE].

\bibitem{Lipari:2008td}
P.~Lipari,
Phys. Rev. 79 (2008) 063001
 and arXiv:0809.0190 [astro-ph].

\bibitem{Ellsworth:1982kv}
R.~Ellsworth, T.~Gaisser, T.~Stanev, and G.~Yodh,
Phys. Rev. D26 (1982) 336.

\bibitem{Baltrusaitis:1985mx}
R.~Baltrusaitis {\it et~al.}  (Fly's Eye Collab.),
Nucl. Instrum. Meth. A240 (1985) 410.

\bibitem{Barbosa:2003dc}
H.~Barbosa, F.~Catalani, J.~A. Chinellato, and C.~Dobrigkeit,
Astropart. Phys. 22 (2004) 159--166
 and astro-ph/0310234.

\bibitem{Dawson:2007di}
B.~R. Dawson  (Pierre Auger Collab.),
{Proc. of 30$^{\rm th}$ Int. Cosmic Ray Conf.}, Merida, Mexico  (2007)
 and arXiv:0706.1105 [astro-ph].

\bibitem{Ulrich:2006hb}
R.~Ulrich, J.~Bl{\"u}mer, R.~Engel, F.~Sch{\"u}ssler, and M.~Unger,
{Proc. of XIV ISVHECRI}, Weihai, China  (2006)
 and astro-ph/0612205.

\bibitem{Ulrich:2009xxx}
R.~Ulrich {\it et~al.},
in preparation  (2009) .

\bibitem{Aglietta:2007yx}
J.~Abraham {\it et~al.}  (Pierre Auger Collab.),
Astropart. Phys. 29 (2008) 243--256
 and arXiv:0712.1147 [astro-ph].

\bibitem{AlvarezMuniz:2004bx}
J.~Alvarez-Muniz, R.~Engel, T.~Gaisser, J.~Ortiz, and T.~Stanev,
Phys. Rev. D69 (2004) 103003
 and astro-ph/0402092.

\bibitem{Block:2007rq}
M.~Block,
Phys. Rev. D76 (2007) 111503
 and arXiv:0705.3037 [hep-ph].

\end{thebibliography}

\end{document}